\documentclass{optica-article}

\journal{opticajournal} 
\articletype{Research Article}

\usepackage{amsmath,amsfonts}
\usepackage{algorithmic}
\usepackage{algorithm}
\usepackage{array}
\usepackage{textcomp}
\usepackage{stfloats}
\usepackage{url}
\usepackage{verbatim}
\usepackage{graphicx}
\usepackage{color}
\usepackage{hyperref}
\usepackage{enumitem}
\usepackage{bm}
\usepackage{booktabs}
\usepackage{array}
\usepackage{subcaption}
\usepackage{media9}
\usepackage{caption}

\begin{document}

\title{EDoF-NeRF:~extended depth-of-field \\ neural radiance fields using \\ a coded aperture camera}

\author{Yoshiyuki Shirasaki\authormark{1} and Ryoichi Horisaki\authormark{1,*}\,\href{https://orcid.org/0000-0002-2280-5921}{\includegraphics[height=1em]{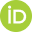}}}

\address{\authormark{1}Department of Information Physics and Computing, Graduate School of Information Science and Technology, The University of Tokyo, 7-3-1 Hongo, Bunkyo-ku, Tokyo 113-8656, Japan}

\email{\authormark{*}horisaki@g.ecc.u-tokyo.ac.jp}

\begin{abstract*}
We propose a method for extending the depth-of-field~(DoF) to construct high-fidelity neural radiance fields~(NeRF)---an emerging technique for rendering photorealistic novel views from a dataset of images captured at different viewpoints, based on implicit neural representations.
The trade-off between DoF and light quantity is inherent not only in conventional cameras but also in NeRF, since the datasets used by NeRF are captured by these cameras.
To address this issue, we introduce a coded aperture placed at the camera pupil, preserving spatial frequency components under defocused conditions.
We develop a camera model incorporating coded apertures into NeRF, allowing direct input of coded images and enabling the generation of novel views with an extended DoF.
We validate the proposed method, termed extended DoF-NeRF~(EDoF-NeRF), through simulations and experiments, demonstrating its superior performance compared to conventional aperture cameras.
\end{abstract*}


\section{Introduction}
Neural radiance fields~(NeRF) have emerged as a powerful framework for synthesizing photorealistic novel views from a sparse set of input images, leveraging implicit neural scene representations~\cite{mildenhall2021NeRF}.
This capability has enabled a wide range of applications, including virtual or augmented reality~\cite{li2022Rt, zou2024ARthroNeRF}, three-dimensional scanning for digital archives~\cite{croce2023Neural, mazzacca2023NeRF}, and autonomous driving~\cite{cao2024Lightning, shen2024Driveenv}.
However, as with other novel-view synthesis techniques, NeRF's practical performance is limited by the inherent trade-off between depth-of-field~(DoF) and light quantity in conventional cameras.
For instance, a large camera aperture, which ensures sufficient light under short exposure conditions, introduces defocus blur due to a shallow DoF.
This trade-off is particularly problematic when capturing dynamic scenes.
Consequently, the limited DoF resulting from this inherent trade-off in NeRF input images directly translates into a similarly limited DoF in the synthesized novel views.

To address this limited DoF in NeRF outputs, previous studies have introduced deblurring methods into the NeRF pipeline, such as Deblur-NeRF~\cite{ma2022Deblur}, aperture rendering~(AR)-NeRF~\cite{kaneko2022ar}, and DoF-NeRF~\cite{wu2022DoF}.
These approaches incorporate defocus-aware rendering models into NeRF, generating deblurred novel views from defocused inputs.
However, defocus blur is inherently an irreversible process due to the loss of high-frequency spatial information~\cite{goodman2005Introduction}.
Consequently, the deblurring performance of these defocus-aware NeRF methods, which rely on conventional cameras, remains insufficient---particularly when capturing input images under shallow-DoF conditions.

In this study, we tackle the issue of limited DoF in NeRF by introducing a coded aperture into the camera used for capturing NeRF inputs.
The coded aperture is a fundamental technique in computational imaging and has been widely applied across various domains~\cite{gehm2007single, wagadarikar2009video, llull2013coded, miyakawa2014coded, bacca2020coupled, ge2023coded, wang2025dual}.
It mitigates the loss of spatial frequency information caused by defocus blur~\cite{ables1968Fourier, dicke1968Scatter, fenimore1978Coded}.
In particular, coded apertures have been used to extend DoF and improve range accuracy in depth-from-defocus methods and related applications~\cite{levin2007Image, zhou2009Coded, horisaki2020Deeply, silva2025toward}.
We develop a defocus rendering model incorporating the coded aperture into the NeRF pipeline.
As a result, the outputs generated by our NeRF exhibit an extended DoF derived from the coded input images.
We validate the proposed method---extended DoF-NeRF~(EDoF-NeRF)---through numerical and optical experiments.
EDoF-NeRF is particularly valuable for novel-view synthesis under conditions of insufficient lighting and/or dynamic scenes, where the trade-off between DoF and light quantity is critical.

\section{Related works}
Several defocus-aware rendering methods related to EDoF-NeRF have recently been proposed to address the limited DoF problem in NeRF:
\begin{itemize}[leftmargin=*]
\item Deblur-NeRF~\cite{ma2022Deblur}:
Removes motion and defocus blur from NeRF inputs by integrating deformable sparse kernels into the NeRF pipeline.
It explicitly models complex blur effects, enhancing image sharpness.
\item AR-NeRF~\cite{kaneko2022ar}:
Learns scene depth and aperture-induced defocus blur simultaneously from natural image datasets in a self-supervised manner.
It enables joint inference of depth maps and realistic DoF rendering.
\item DoF-NeRF~\cite{wu2022DoF}:
Provides explicit control of DoF effects within NeRF through a physically based rendering model.
It enables rendering novel views with adjustable focus settings, including large-DoF conditions.
\end{itemize}
These approaches assume conventional circular apertures typically used in standard cameras.
Therefore, achieving greater DoF extension remains challenging due to inherent optical constraints.
The proposed EDoF-NeRF leverages the advantages of coded apertures, enabling a further extension of DoF compared to these previous methods.

\section{Rendering models}
\label{sec:rendering_models}
\subsection{Pinhole camera}
\begin{figure}[t]
\centering
\includegraphics[scale=0.65]{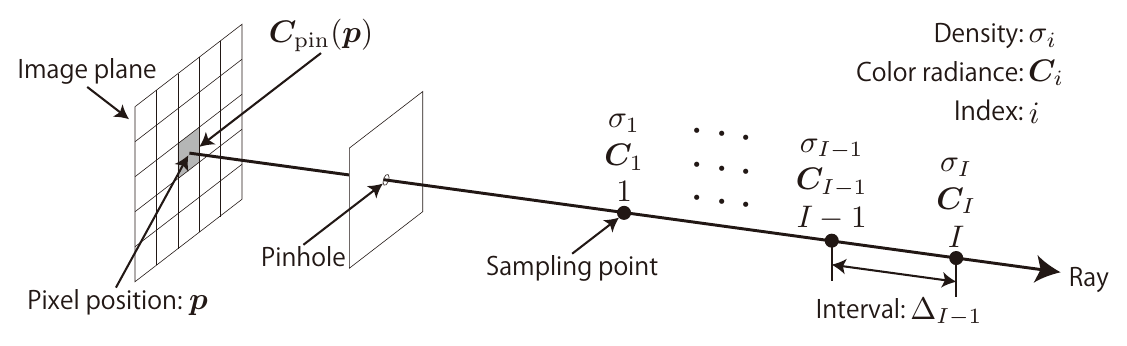}
\caption{Pinhole camera model.}
\label{fig:model_pinhole}
\end{figure}

Original NeRFs employ a pinhole camera model to synthesize the color~$\bm{C}_\text{pin}(\bm{p})$ at the pixel position~$\bm{p}$ on the image plane, as illustrated in Fig.~\ref{fig:model_pinhole}~\cite{mildenhall2021NeRF}.
Along the ray connecting the pinhole and the pixel position~$\bm{p}$, the color radiance~$\bm{C}_i(\bm{p})$ at the sampling point indexed by $i$, with density~$\sigma_i(\bm{p})$, is integrated using the opacity~$O_i(\bm{p})$, which determines the local contribution by attenuating signals from upstream samples, and the transmittance~$T_i(\bm{p})$, which represents attenuation caused by downstream samples, as
\begin{equation}
\bm{C}_\text{pin}(\bm{p}) = \sum_{i=1}^{I(\bm{p})} T_i(\bm{p})O_i(\bm{p})\bm{C}_i(\bm{p}),
\label{eq:pinhole}
\end{equation}
where $I(\bm{p})$ is the number of sampling points along the ray.
The opacity~$O_i(\bm{p})$ and transmittance~$T_i(\bm{p})$ are defined as
\begin{equation}
O_i(\bm{p}) = 1 - \exp\left(-\sigma_i(\bm{p}) \Delta_i(\bm{p})\right),
\label{eq:attn}
\end{equation}
and
\begin{equation}
T_i(\bm{p}) = \exp\left(-\sum_{j=1}^{i-1}\sigma_j(\bm{p}) \Delta_j(\bm{p})\right),
\label{eq:trns}
\end{equation}
respectively, where $\Delta_i(\bm{p})$ is the interval between sampling points $i$ and $i-1$.
As shown in Eqs.~\eqref{eq:attn} and \eqref{eq:trns}, a larger density~$\sigma_i(\bm{p})$ and a smaller interval~$\Delta_i(\bm{p})$ increase the opacity~$O_i(\bm{p})$ and decrease the transmittance~$T_i(\bm{p})$, and vice versa.

\subsection{Lens camera}
\begin{figure}[t]
\centering
\includegraphics[scale=0.65]{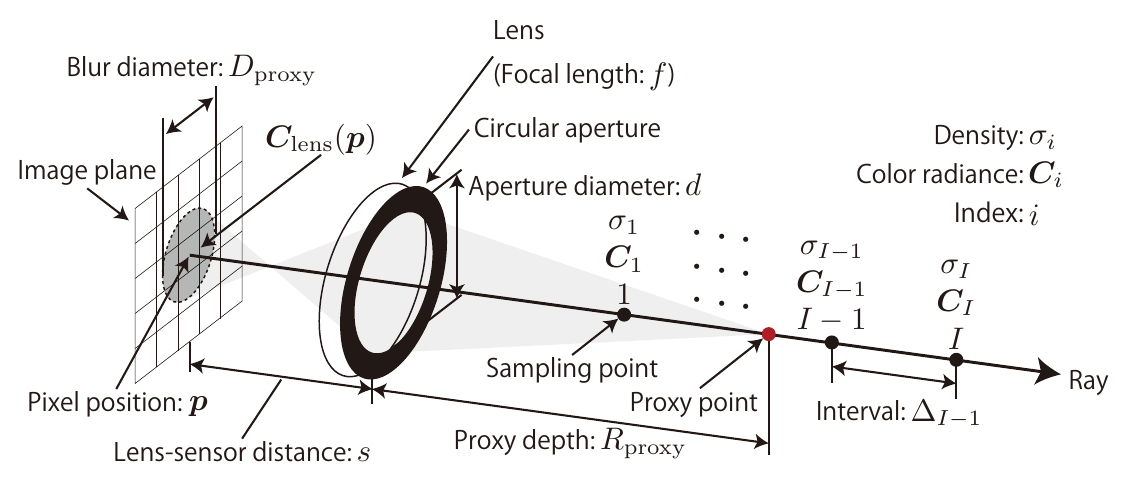}
\caption{Lens camera model.}
\label{fig:model_lens}
\end{figure}

The lens camera model proposed in DoF-NeRF~\cite{wu2022DoF}, as illustrated in Fig.~\ref{fig:model_lens}, simulates the defocus blur introduced by the finite aperture of physical cameras.
This model extends the pinhole formulation by computing the diameter~$D_\text{proxy}(\bm{p})$ of the point spread function~(PSF) that characterizes the blur at a proxy point representing the multiple sampling points along a ray, based on the proxy depth~$R_\text{proxy}(\bm{p})$, defined as the axial distance from the lens to the proxy point.
The proxy depth~$R_\text{proxy}(\bm{p})$ is defined as the weighted average of the axial distances~$r_i(\bm{p})$ from the lens to the sampling points, where the weights correspond to their contributions to the color radiance at the pixel position~$\bm{p}$ on the image plane:
\begin{equation}
R_\text{proxy}(\bm{p}) = \frac{\sum_{i=1}^{I(\bm{p})} T_i(\bm{p})O_i(\bm{p})r_i(\bm{p})}{\sum_{i=1}^{I(\bm{p})} T_i(\bm{p})O_i(\bm{p})}.
\label{eq:proxy_depth}
\end{equation}
The blur diameter~$D_\text{proxy}(\bm{p})$ corresponding to the proxy depth~$R_\text{proxy}(\bm{p})$ is expressed as
\begin{equation}
D_\text{proxy}(\bm{p}) = f d \left| \frac{1}{z} - \frac{1}{R_\text{proxy}(\bm{p})} \right|,
\label{eq:blur_diameter}
\end{equation}
where $f$ is the focal length of the lens, $d$ is the aperture diameter, and $z$ is the camera focusing distance, given by $z = (1/f - 1/s)^{-1}$, with $s$ denoting the lens--sensor distance.
Finally, the color radiance~$\bm{C}_\text{lens}(\bm{p})$ of the lens camera model is obtained by a blur convolution applied to the pinhole color radiance~$\bm{C}_\text{pin}(\bm{p})$:
\begin{equation}
\bm{C}_\text{lens}(\bm{p}) = \sum_{\bm{p}'} \bm{C}_\text{pin}(\bm{p}') P_\text{lens}(\bm{p} - \bm{p}', D_\text{proxy}(\bm{p}')),
\label{eq:lens_rendering}
\end{equation}
where $P_\text{lens}(\bm{p}, D_\text{proxy}(\bm{p}))$ is a position-dependent~(shift-variant) blur PSF defined by a disk function with diameter~$D_\text{proxy}(\bm{p})$ at the pixel position~$\bm{p}$.

\subsection{Coded aperture camera}
\begin{figure}[t]
\centering
\includegraphics[scale=0.65]{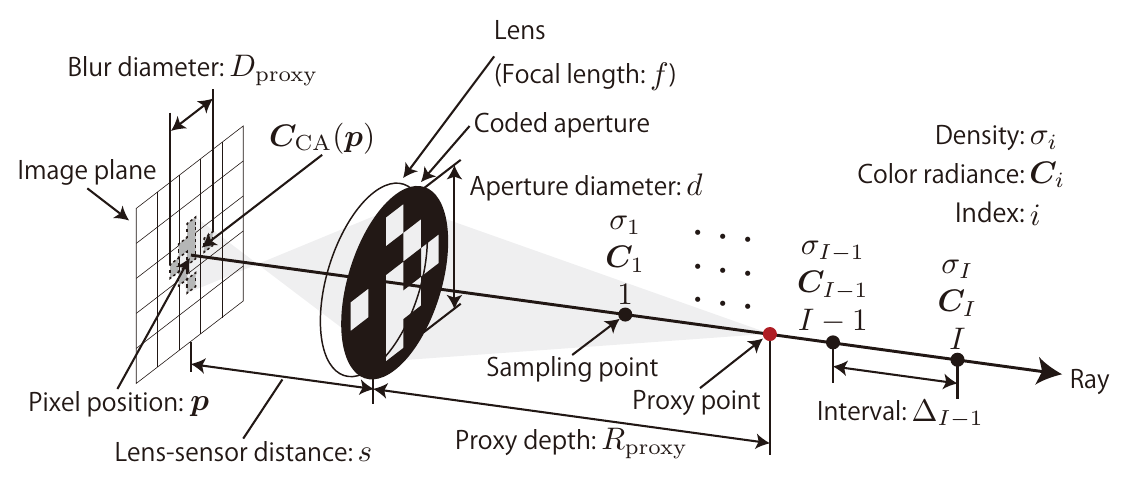}
\caption{Coded aperture camera model.}
\label{fig:model_coded}
\end{figure}

To mitigate the loss of high-frequency spatial information caused by defocus blur in the lens camera model used in DoF-NeRF~\cite{wu2022DoF}, we introduce a coded aperture mask on the pupil plane to extend the DoF in NeRF by spatially controlling light transmission, as illustrated in Fig.~\ref{fig:model_coded}~\cite{ables1968Fourier, dicke1968Scatter, fenimore1978Coded, levin2007Image, zhou2009Coded, horisaki2020Deeply}.
Similar to the lens camera model in Eq.~\eqref{eq:lens_rendering}, the color radiance~$\bm{C}_\text{CA}(\bm{p})$ of the coded aperture camera is obtained by applying a coded-PSF convolution to the pinhole color radiance~$\bm{C}_\text{pin}(\bm{p})$:
\begin{equation}
\bm{C}_\text{CA}(\bm{p}) = \sum_{\bm{p}'} \bm{C}_\text{pin}(\bm{p}') P_\text{CA}(\bm{p} - \bm{p}', D_\text{proxy}(\bm{p}')),
\label{eq:ca_rendering}
\end{equation}
where~$P_\text{CA}(\bm{p}, D_\text{proxy}(\bm{p}))$ is a position-dependent coded PSF
induced by the mask, with diameter~$D_\text{proxy}(\bm{p})$ derived in Eq.~\eqref{eq:blur_diameter}, at the pixel position~$\bm{p}$.

\section{Network architecture}
\subsection{Positional encoding}
\begin{figure}[t]
\centering
\includegraphics[width=0.9\linewidth]{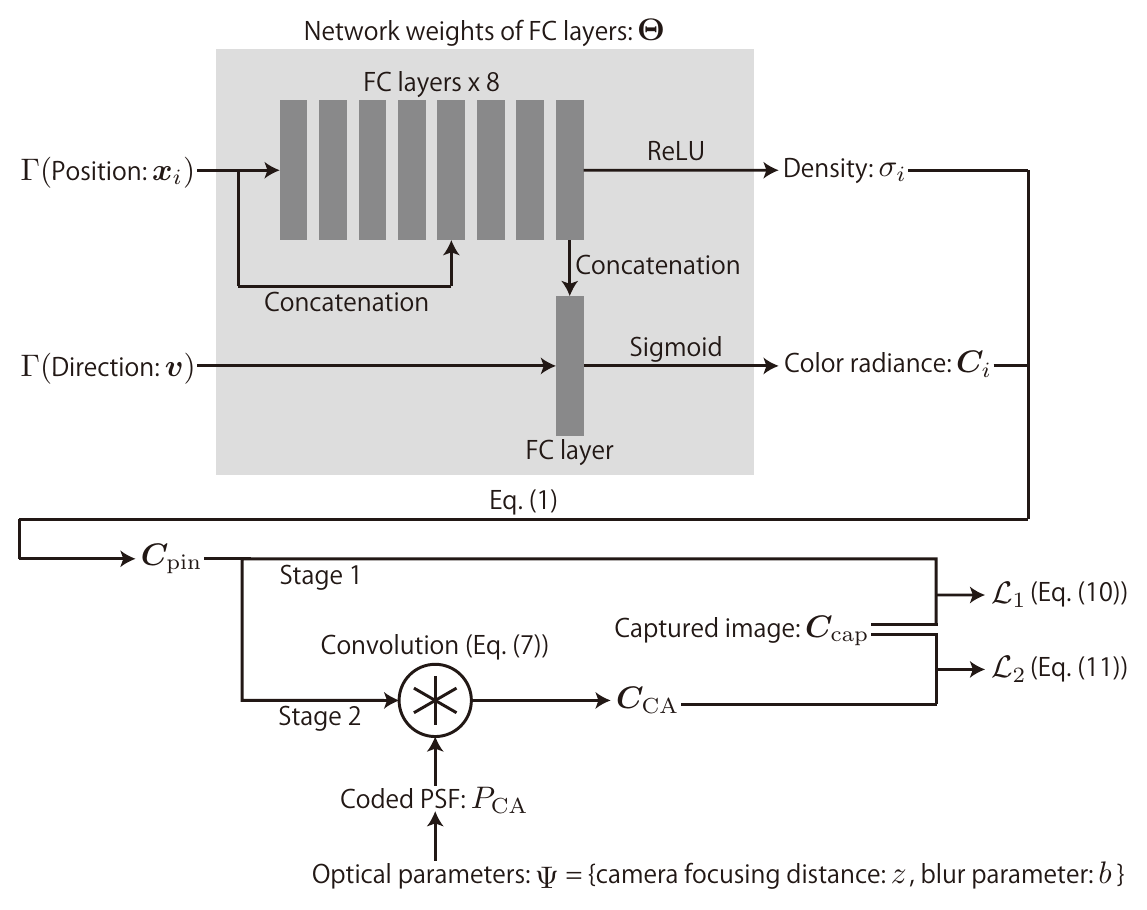}
\caption{Overview of the network architecture and training pipeline. FC~layer:~fully connected layer; ReLU:~rectified linear unit.}
\label{fig:method}
\end{figure}

Following the original NeRF framework, we employ multi-layer perceptrons~(MLPs) to predict the color radiance~$\bm{C}_i(\bm{p})$ and the density~$\sigma_i(\bm{p})$ at the $i$-th sampling point along the ray corresponding to the pixel position~$\bm{p}$, as shown in Fig.~\ref{fig:method}~\cite{mildenhall2021NeRF}.
The three-dimensional position vector~$\bm{x}_i(\bm{p}) = (x_i(\bm{p}), y_i(\bm{p}), z_i(\bm{p}))$ of the sampling point and the two-dimensional ray direction vector~$\bm{v}(\bm{p}) = (\theta(\bm{p}), \phi(\bm{p}))$ are provided to the MLPs via positional encoding~$\Gamma(\bullet)$ to capture high-frequency information of the scene, defined as
\begin{equation}
\Gamma(\bm{u}) = \left( \sin(2^0 \pi \bm{u}), \cos(2^0 \pi \bm{u}), \dots, \sin(2^{N-1} \pi \bm{u}), \cos(2^{N-1} \pi \bm{u}) \right),
\label{eq:positional_encoding}
\end{equation}
where $\bm{u}$ denotes an input vector corresponding to either the position vector~$\bm{x}_i(\bm{p})$ or the direction vector~$\bm{v}(\bm{p})$, and $N$ is the number of frequency bands.
The positional encoding~$\Gamma(\bullet)$ is applied element-wise to each component of the input vector~$\bm{u}$.

\subsection{MLP structure}
The MLP architecture comprises two branches, as illustrated in Fig.~\ref{fig:method}~\cite{mildenhall2021NeRF}.
The first branch predicts the density~$\sigma_i(\bm{p})$ from the position vector~$\bm{x}_i(\bm{p})$ using positional encoding~$\Gamma(\bullet)$ defined in Eq.~\eqref{eq:positional_encoding}.
The second branch predicts the color radiance~$\bm{C}_i(\bm{p})$ from both the position vector~$\bm{x}_i(\bm{p})$ and the direction vector~$\bm{v}(\bm{p})$, also using positional encoding~$\Gamma(\bullet)$.
This architecture ensures that the density~$\sigma_i(\bm{p})$ depends only on the position vector~$\bm{x}_i(\bm{p})$, representing a consistent scene geometry, while the color radiance~$\bm{C}_i(\bm{p})$ varies with the direction vector~$\bm{v}(\bm{p})$ to model view-dependent lighting effects such as specular reflections.

The first branch contains eight fully connected~(FC) layers, each with 256~channels and employing the rectified linear unit~(ReLU) activation function.
To mitigate the vanishing gradient problem and preserve high-frequency information, the positional encoding output~$\Gamma(\bm{x}_i(\bm{p}))$ is directly concatenated with the input to the fifth layer via a skip connection.
Subsequently, the second branch takes the output vector from the eighth layer of the first branch, which captures the scene geometry, and concatenates it with the positional encoding output~$\Gamma(\bm{v}(\bm{p}))$.
The resulting vector is then processed through a single FC~layer with 128~channels and a sigmoid activation function to constrain the color radiance~$\bm{C}_i(\bm{p})$ to the range~$[0, 1]$.

\section{Training}
\subsection{Patch sampling}
\label{subsec:patch}
In the original NeRF, the MLPs are trained using random ray sampling, where the pixel position~$\bm{p}$ is randomly selected and $\bm{C}_\text{pin}$ is computed based on Eq.~\eqref{eq:pinhole}~\cite{mildenhall2021NeRF}.
In contrast, DoF-NeRF as well as EDoF-NeRF employ patch sampling for convolution-based rendering~\cite{wu2022DoF}.
In the patch-sampling approach, a patch is selected and rendered to obtain $\bm{C}_\text{pin}$ within the patch, which is then convolved with the PSFs as shown in Eqs.~\eqref{eq:lens_rendering} and \eqref{eq:ca_rendering}.

In this study, we introduce a margin around each square patch to accurately compute the loss function described in the next subsection by suppressing the influence of boundary effects arising from convolution operations, as
\begin{equation}
S_\text{patch} = S_\text{loss} + M,
\label{eq:patch_size}
\end{equation}
where $S_\text{patch}$ is the side length of the square patch, $S_\text{loss}$ is the side length of the square region used to compute the loss function, and $M$ is the width of the margin.
Note that the margin~$M$ is set to 0 in DoF-NeRF.
The maximum margin~$M$ corresponds to the maximum value of $D_\text{proxy}$; however, a large $M$ results in either high computational cost~(large $S_\text{patch}$) or low training throughput~(small $S_\text{loss}$).
Therefore, in EDoF-NeRF, we empirically set the margin~$M$ in each experiment by balancing this trade-off.
During training based on patch sampling in EDoF-NeRF, the patch interval is defined as $\lfloor S_\text{loss} / 3 \rfloor$, following DoF-NeRF~\cite{wu2022DoF}.

\subsection{Two-stage training}
\label{subsec:two_stage}
Direct optimization of the coded aperture camera model defined in Eq.~\eqref{eq:ca_rendering} is unstable, as it involves not only optimizing the weight parameters~$\bm{\Theta}$ of the MLPs shown in Fig.~\ref{fig:method}, but also estimating the optical parameters~$\bm{\Psi}_m$ for the $m$-th captured image.
These optical parameters consist of the camera focusing distance~$z_m$ and a blur parameter~$b_m$, defined as the product of the focal length~$f$ and the aperture diameter~$d$, which appear in Eq.~\eqref{eq:blur_diameter}.

To achieve stable optimization, we adopt the following two-stage training strategy, as illustrated in Fig.~\ref{fig:method}~\cite{wu2022DoF}:
\begin{description}
\item[Stage~1:]
We optimize only the network weights~$\bm{\Theta}$ using the pinhole camera model in Eq.~\eqref{eq:pinhole}.
Although this model does not correctly account for the coded PSFs, it enables stable and rapid estimation of the coarse scene geometry.
In this stage, we employ random ray sampling over the entire image, where no convolution process is involved~\cite{mildenhall2021NeRF}.
The loss function for Stage~1 is defined as the mean squared error~(MSE) between the color radiance~$\bm{C}_\text{pin}(\bm{p}_m ; \bm{\Theta})$ of the $m$-th image rendered by the pinhole camera model and the color radiance~$\bm{C}_\text{cap}(\bm{p}_m)$ of the corresponding $m$-th captured image:
\begin{equation}
\mathcal{L}_1(\bm{\Theta}) =
\frac{1}{\sum_{m \in \mathcal{M}_1} |\mathcal{R}_m|}\sum_{m \in \mathcal{M}_1}~
\sum_{\bm{p}_m \in \mathcal{R}_m}
\left\|
\bm{C}_{\text{pin}}(\bm{p}_m ; \bm{\Theta})-\bm{C}_\text{cap}(\bm{p}_m)
\right\|_2^2,
\label{eq:loss_1}
\end{equation}
where $\mathcal{M}_1$ denotes a randomly sampled set of image indices~$m$, and $\mathcal{R}_m$ denotes a randomly sampled set of pixel positions~$\bm{p}_m$ on the $m$-th image, used for stochastic mini-batch training based on random ray sampling.
\item[Stage~2:]
We then use the coded aperture camera model and adopt patch sampling with the margin discussed in Subsection~\ref{subsec:patch} to incorporate the coded-PSF convolution, thereby fine-tuning the network to recover high-frequency details of the scene.
In this stage, we jointly optimize the network weights~$\bm{\Theta}$ and the optical parameters~$\bm{\Psi}_m = \{z_m, b_m\}$ for each captured image.
The loss function for Stage~2 is defined as the MSE between the color radiance~$\bm{C}_\text{CA}(\bm{p}_m ; \bm{\Theta}, \bm{\Psi}_m)$ of the $m$-th image rendered by the coded aperture camera model and the color radiance~$\bm{C}_\text{cap}(\bm{p}_m)$ of the corresponding $m$-th captured image:
\begin{equation}
\mathcal{L}_2(\bm{\Theta}, \bm{\Psi}_{m \in \mathcal{M}_2}) =
\frac{1}{\sum_{m \in \mathcal{M}_2} |\mathcal{P}_m|}\sum_{m \in \mathcal{M}_2}~
\sum_{\bm{p}_m \in \mathcal{P}_m}
\left\|
\bm{C}_\text{CA}(\bm{p}_m ; \bm{\Theta}, \bm{\Psi}_m)-\bm{C}_\text{cap}(\bm{p}_m)
\right\|_2^2,
\label{eq:loss_2}
\end{equation}
where $\mathcal{M}_2$ denotes a singleton set containing a single randomly selected image index~$m$, and $\mathcal{P}_m$ denotes a set of pixel positions~$\bm{p}_m$ inside the patch on the $m$-th image, used for stochastic mini-batch training based on patch sampling.
Note that the color radiance~$\bm{C}_\text{CA}$ is rendered over the extended patch including the margin, whereas the loss is computed only over the patch~$\mathcal{P}_m$, which excludes the margin, as described in Subsection~\ref{subsec:patch}.
\end{description}

The key difference between Stage~1 and Stage~2 lies in both the forward rendering model and the optimization objective: Stage~1 performs geometry initialization under a simplified pinhole formulation without convolution, whereas Stage~2 adopts the full coded aperture model with convolution and jointly refines both the scene geometry and the optical parameters.
In each stage, the loss functions in Eqs.~\eqref{eq:loss_1} and \eqref{eq:loss_2} are minimized using gradient descent with the Adam optimizer~\cite{kingma2014adam}.

\section{Demonstration}
We demonstrated the proposed EDoF-NeRF through both numerical evaluation and optical validation, comparing it with the original NeRF and DoF-NeRF in the following subsections.
For the numerical evaluation, we used a publicly available NeRF dataset, whereas for the optical validation, we employed a commercial camera equipped with a laser-cut coded aperture.

\subsection{Numerical demonstration}
\begin{figure}[tbp]
  \centering
  
  \begin{subfigure}[t]{0.32\linewidth}
    \centering
    \includegraphics[height=4.2cm]{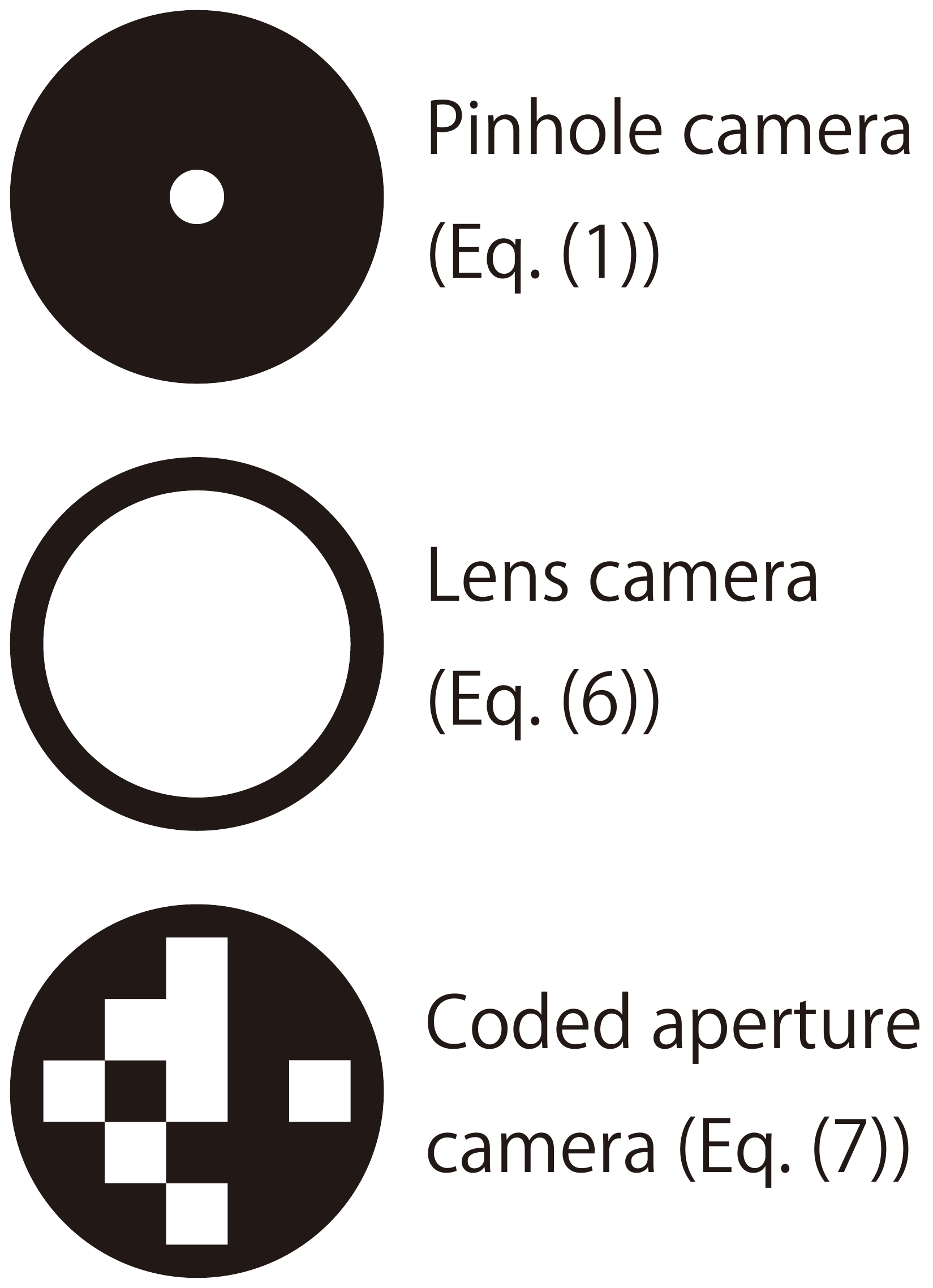}
    \vspace{-2mm}
  \end{subfigure}
  \hfill
  \begin{subfigure}[t]{0.32\linewidth}
    \centering
    \includegraphics[height=4.5cm]{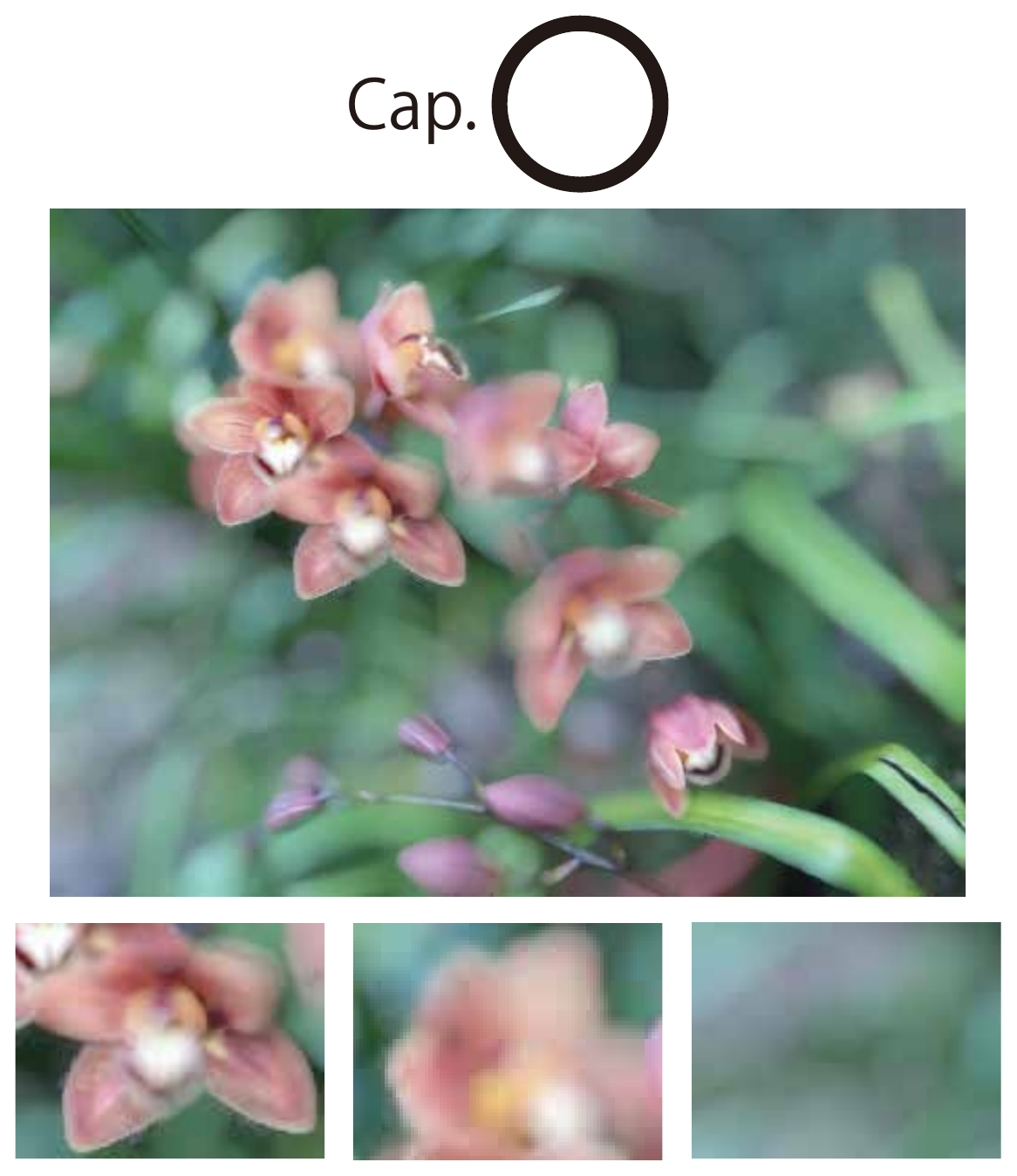}
    \vspace{-2mm}
    \caption*{(b)}
  \end{subfigure}
  \hfill
  \begin{subfigure}[t]{0.32\linewidth}
    \centering
    \includegraphics[height=4.5cm]{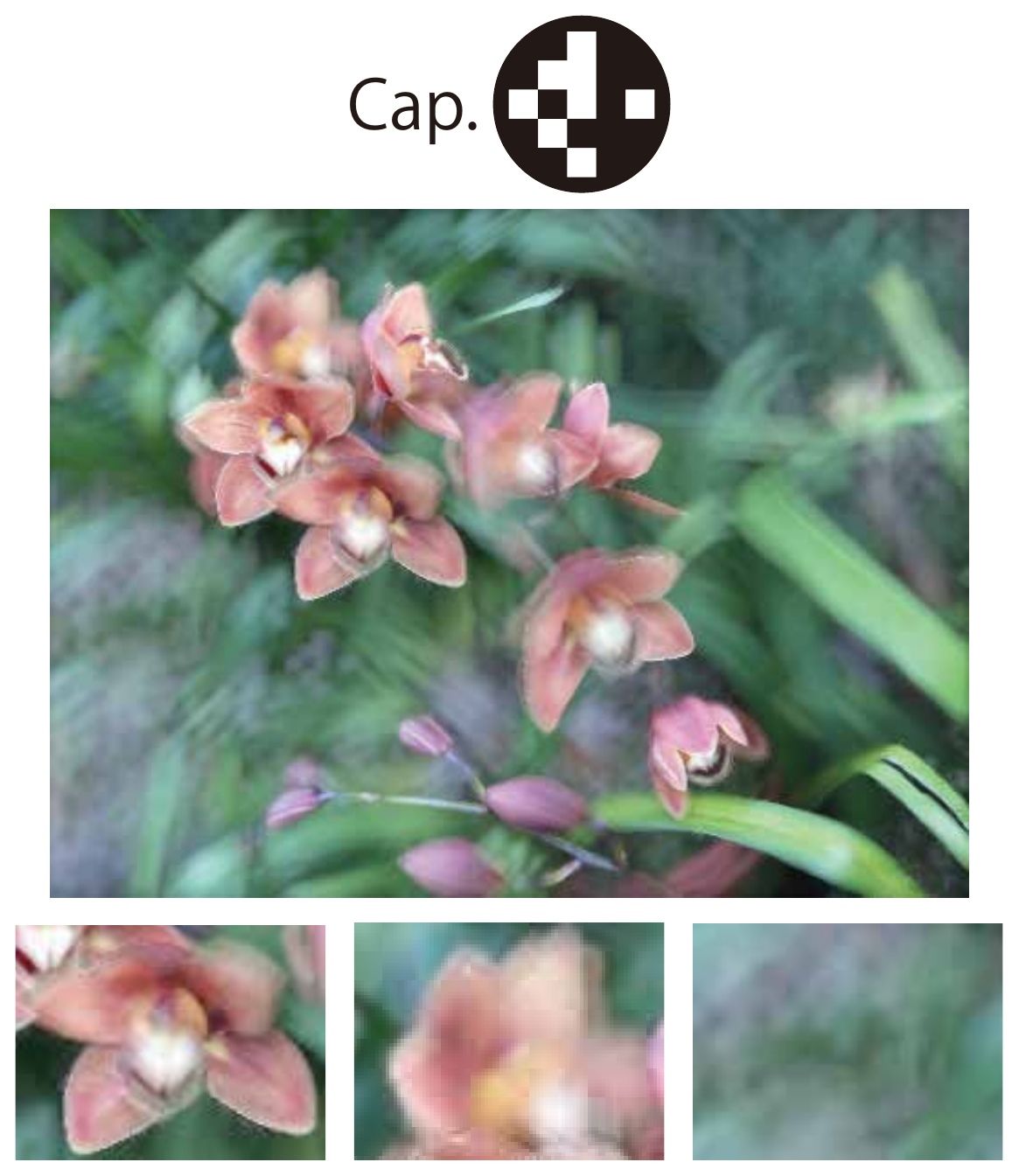}
    \vspace{-2mm}
    \caption*{(c)}
  \end{subfigure}

  \vspace{4mm}

  \begin{subfigure}[t]{0.32\linewidth}
    \centering
    \includegraphics[height=4.5cm]{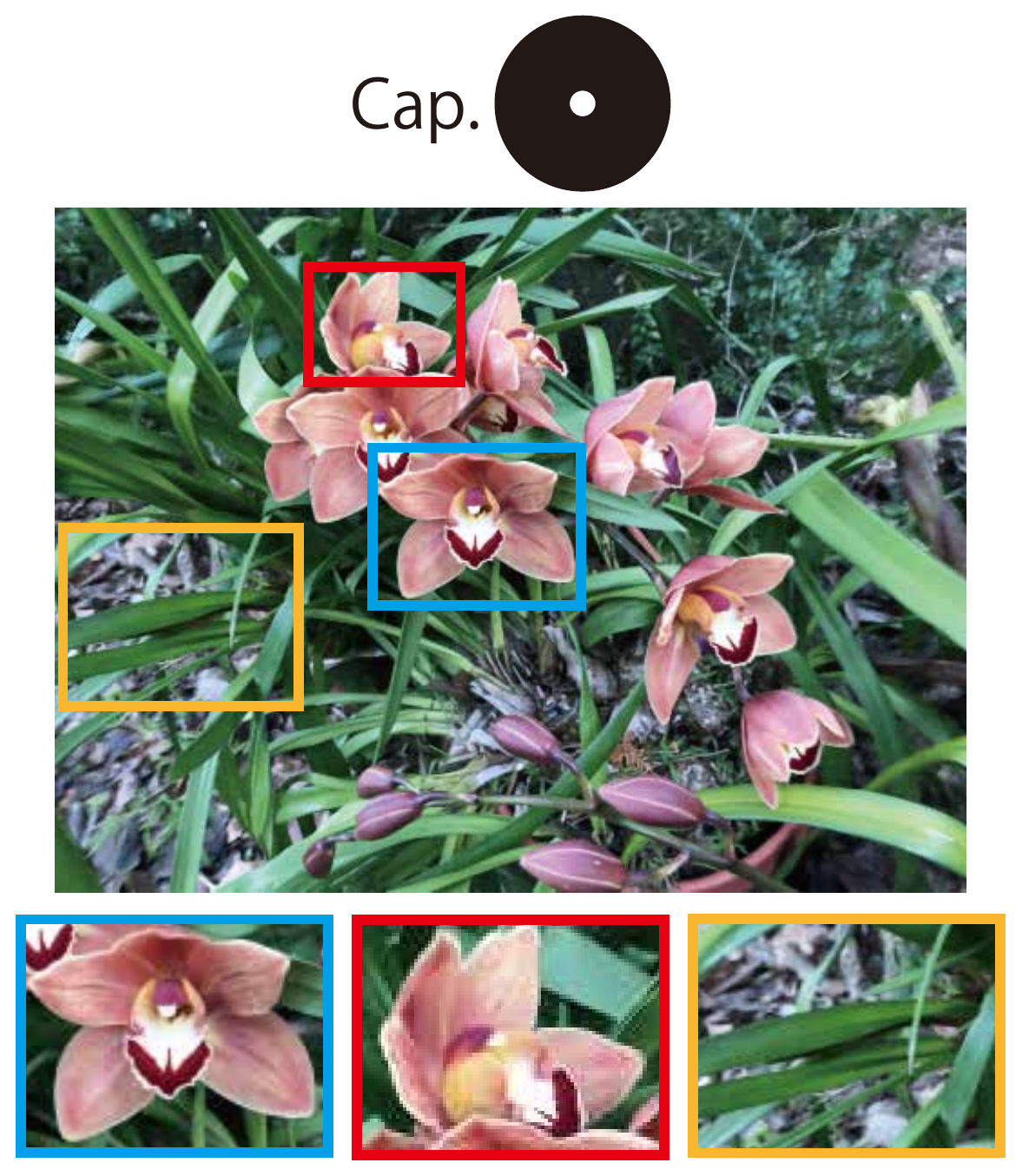}
    \vspace{-2mm}
    \caption*{(a)}
  \end{subfigure}
  \hfill
  \begin{subfigure}[t]{0.32\linewidth}
    \centering
    \includegraphics[height=4.5cm]{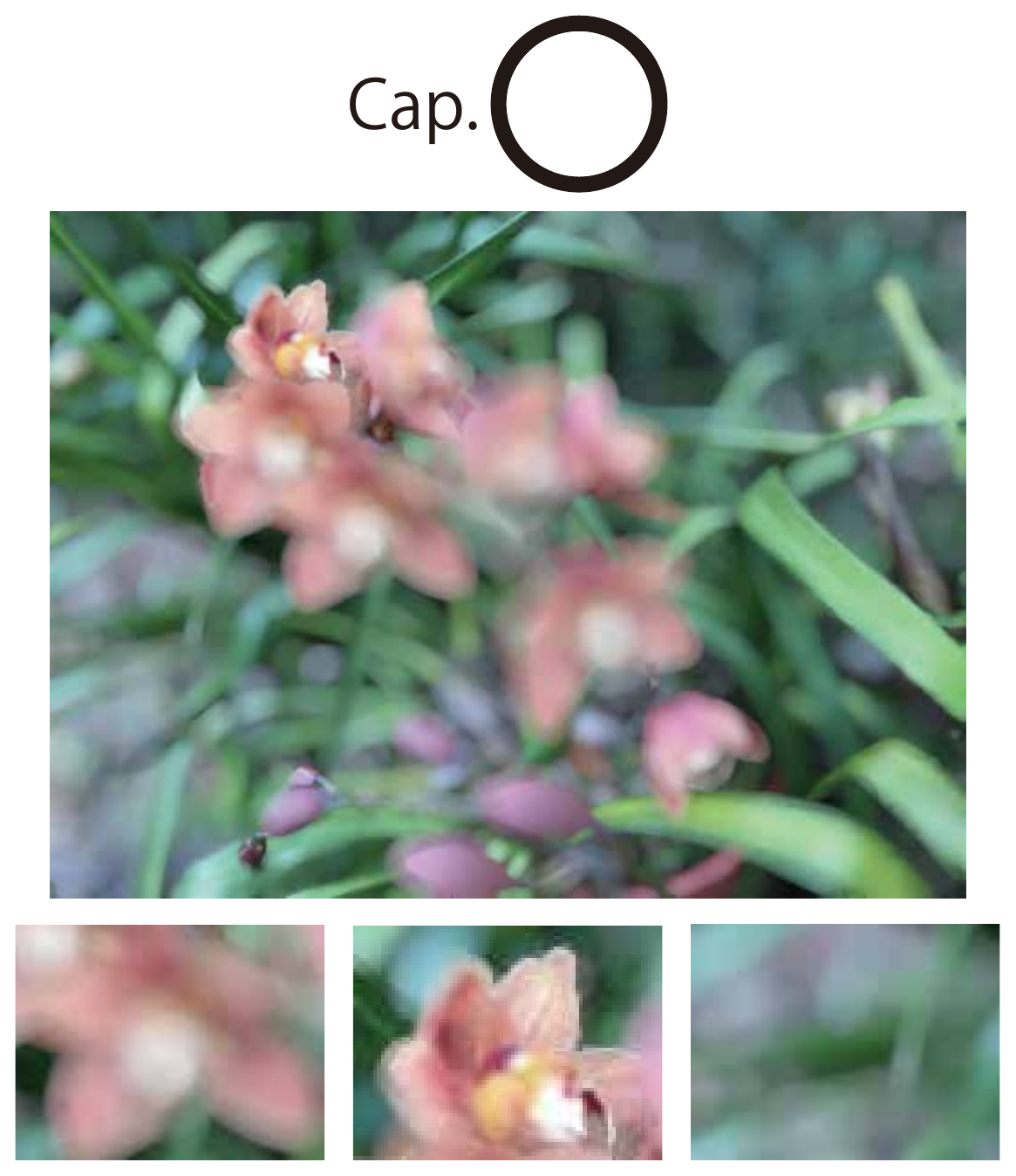}
    \vspace{-2mm}
    \caption*{(d)}
  \end{subfigure}
  \hfill
  \begin{subfigure}[t]{0.32\linewidth}
    \centering
    \includegraphics[height=4.5cm]{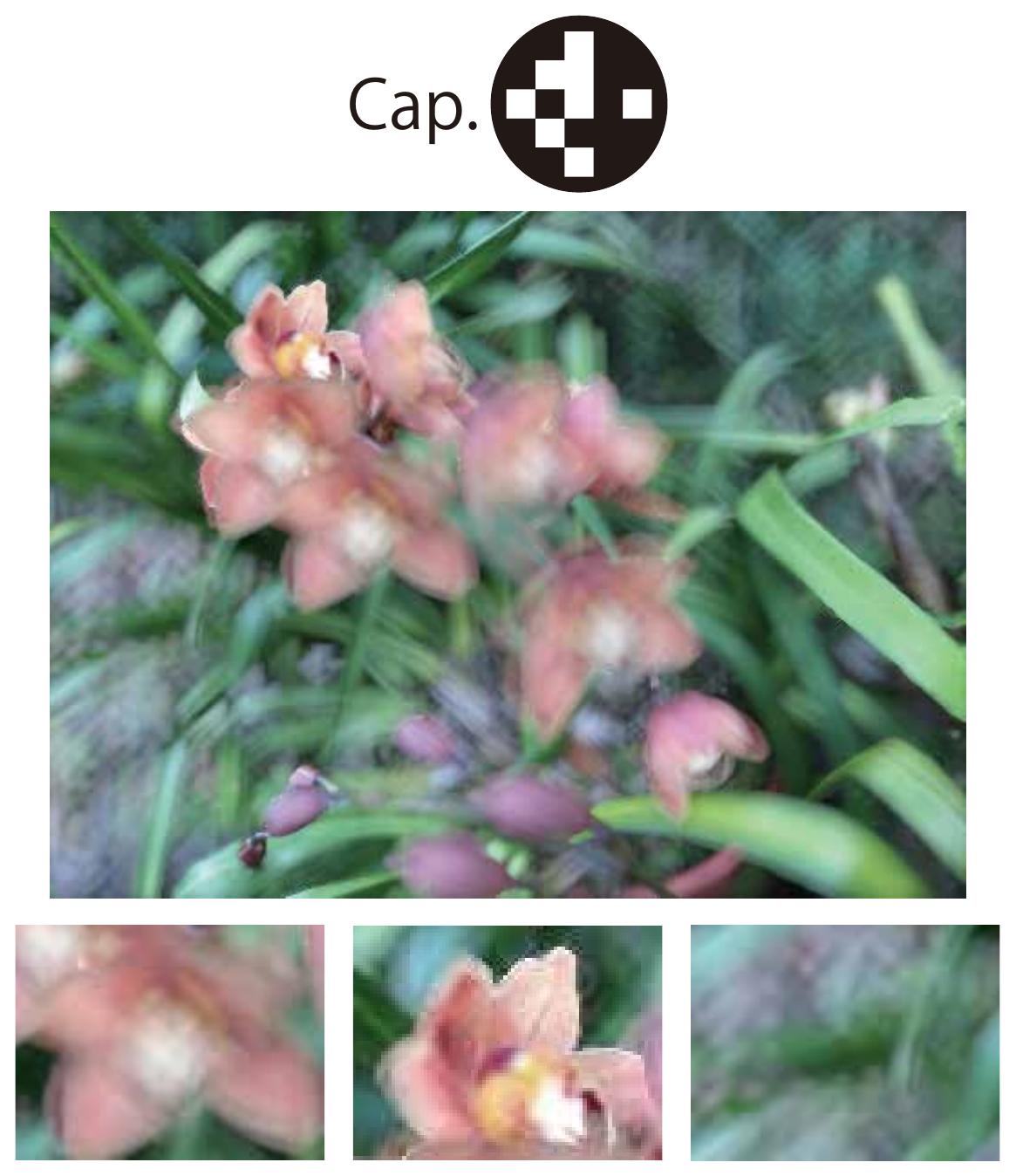}
    \vspace{-2mm}
    \caption*{(e)}
  \end{subfigure}

  \vspace{6mm}

  \begin{subfigure}[t]{0.32\linewidth}
    \centering
    \includegraphics[height=4.5cm]{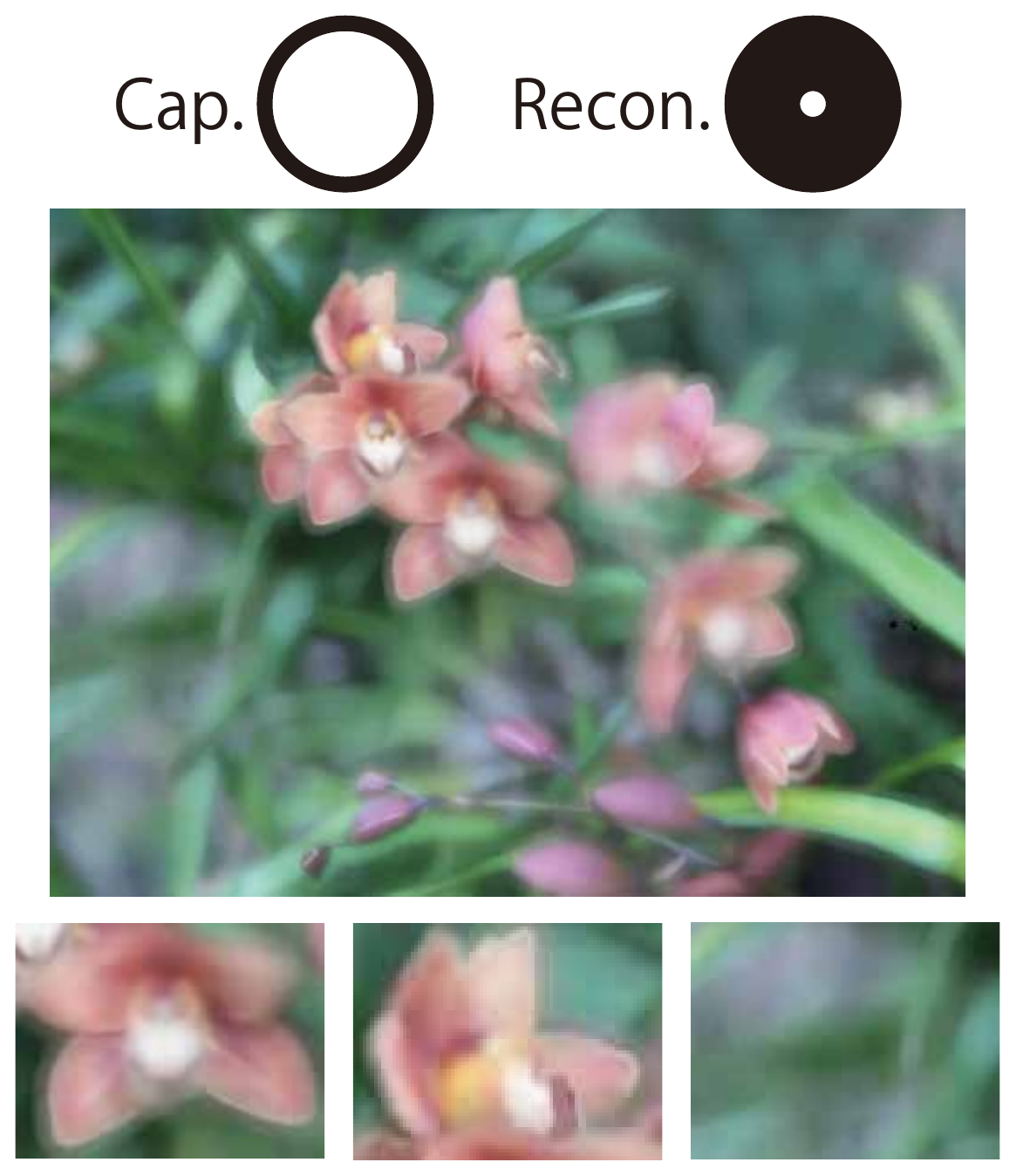}
    \vspace{-2mm}
    \caption*{(f)}
  \end{subfigure}
  \hfill
  \begin{subfigure}[t]{0.32\linewidth}
    \centering
    \includegraphics[height=4.5cm]{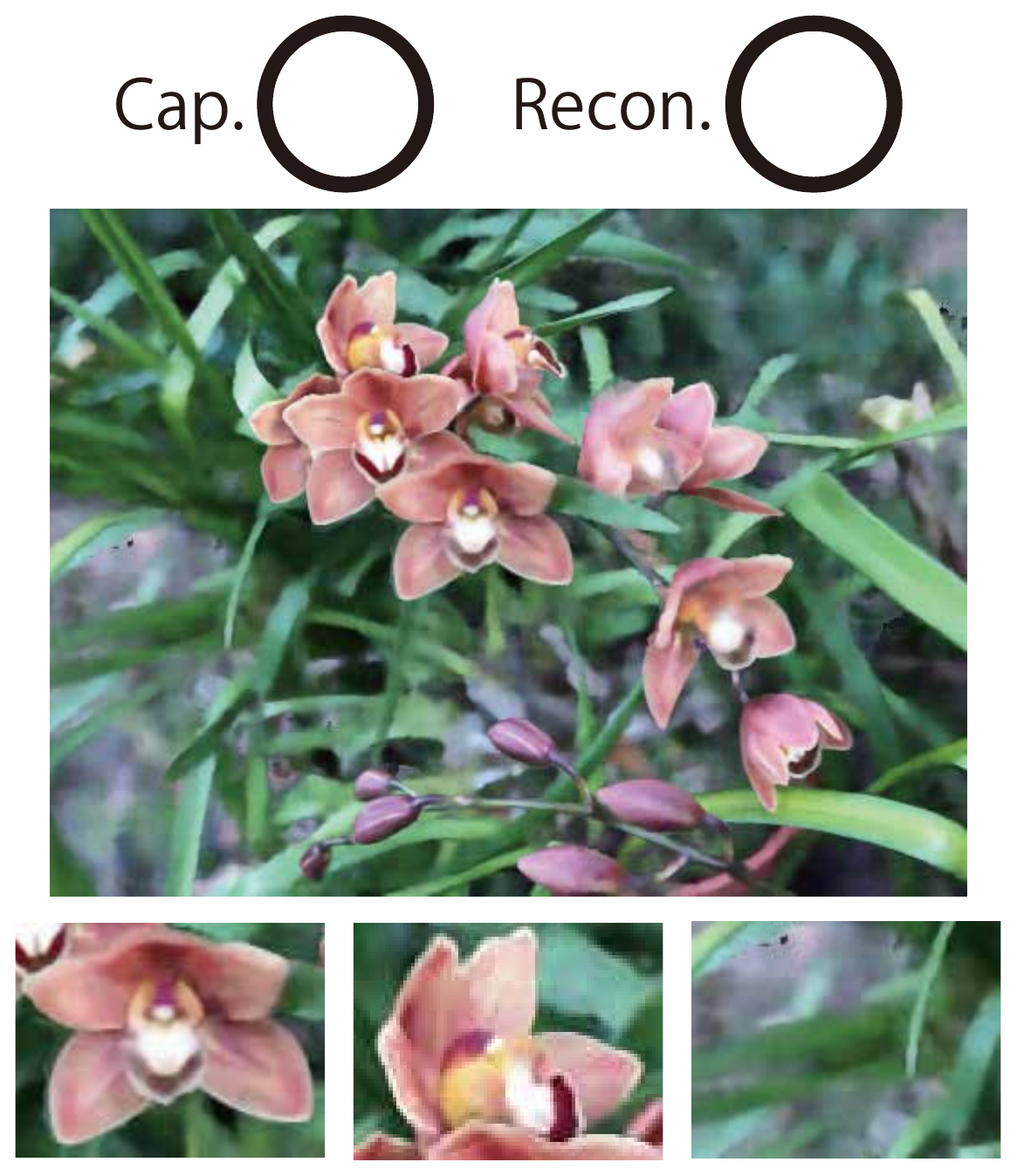}
    \vspace{-2mm}
    \caption*{(g)}
  \end{subfigure}
  \hfill
  \begin{subfigure}[t]{0.32\linewidth}
    \centering
    \includegraphics[height=4.5cm]{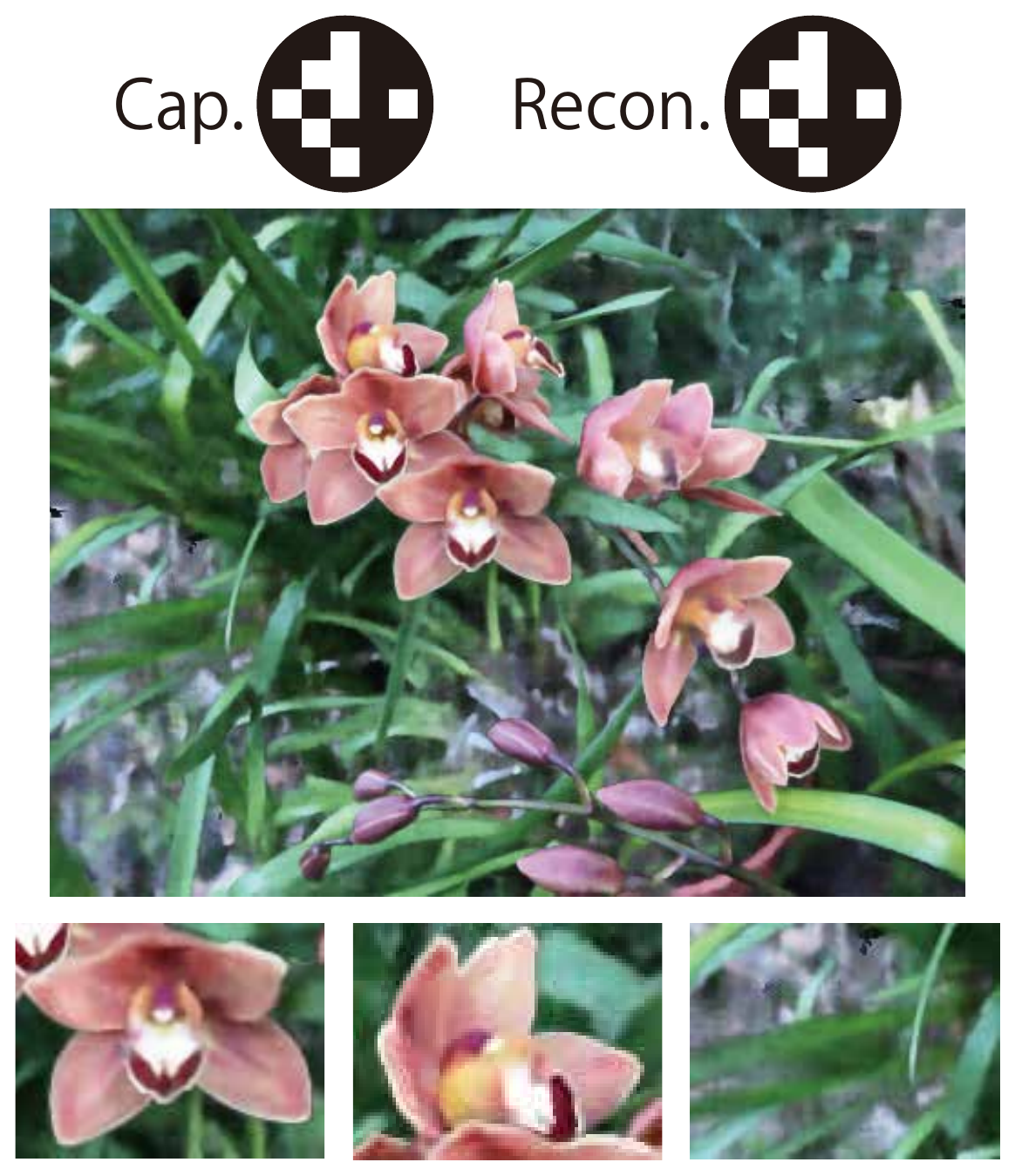}
    \vspace{-2mm}
    \caption*{(h)}
  \end{subfigure}

  \vspace{2mm}
  \caption{
  Numerical demonstration results.
  (a)~Reference novel-view image generated by the pinhole camera model.
  Foreground-focused captured images generated by the (b)~lens and (c)~coded aperture camera models.
  Background-focused captured images generated by the (d)~lens and (e)~coded aperture camera models.
  Reconstruction results from (d) using the (f)~pinhole and (g)~lens camera models.
  (h)~Reconstruction result from (e) using the coded aperture camera model.
  At the bottom of each subfigure, zoomed-in views of objects at different depths are shown; the corresponding regions are indicated by colored boxes in (a).
  The pupil illustrations used in the simulations are shown as legends in the top-left corner.
  Cap.:~capture; Recon.:~reconstruction.
  }
  \label{fig:num_rgb}
\end{figure}

To conduct the numerical demonstration, we prepared a reference NeRF model trained on all-in-focus images of the \textit{Orchids} natural scene from the original NeRF dataset~\cite{mildenhall2021NeRF}.  
As shown in Fig.~\ref{fig:num_rgb}(a), this reference NeRF synthesized all-in-focus~(large DoF) novel-view images formed from the ideal color radiance~$\bm{C}_\text{pin}$ under the pinhole camera model described in Eq.~\eqref{eq:pinhole} and Fig.~\ref{fig:method}.  
By appending the blur convolution process described in Eq.~\eqref{eq:lens_rendering} to the reference NeRF, novel-view images rendered through the lens camera model were synthesized, corresponding to the captured images~$\bm{C}_\text{cap}$ in Fig.~\ref{fig:method}, as shown in Fig.~\ref{fig:num_rgb}(b).  
Similarly, novel-view images rendered through the coded aperture camera model using Eq.~\eqref{eq:ca_rendering} were generated, corresponding to the captured images~$\bm{C}_\text{cap}$, as presented in Fig.~\ref{fig:num_rgb}(c), where the mask pattern on the coded aperture, shown in the legends of Fig.~\ref{fig:num_rgb}, was empirically selected in this study.

In this simulation, we considered 24 viewpoints for novel-view synthesis.  
Following the procedure in DoF-NeRF~\cite{wu2022DoF}, at each viewpoint, we generated images using two camera focusing distances in Eq.~\eqref{eq:blur_diameter}: $z_m = 0.4$ to focus on the foreground object and $z_m = 0.5$ to focus on the background object.
The blur parameter~$b_m~(= fd~\text{in Eq.~\eqref{eq:blur_diameter}})$ was set to $30$ under all these conditions.  
All novel-view images produced by the lens camera model and the coded aperture camera model under these conditions were synthesized using the process described in the preceding paragraph, corresponding to the captured images~$\bm{C}_\text{cap}$.    
The representative captured images~$\bm{C}_\text{cap}$ under foreground and background focus produced by the lens camera model are shown in Figs.~\ref{fig:num_rgb}(b) and \ref{fig:num_rgb}(d), respectively, while those produced by the coded aperture camera model are shown in Figs.~\ref{fig:num_rgb}(c) and \ref{fig:num_rgb}(e).

The first numerical result corresponded to NeRF, in which the captured images, $\bm{C}_\text{cap}$, generated using the lens camera model were used to train the network under the pinhole camera model.
In this case, strong defocus blur appeared in the novel-view image, as shown in Fig.~\ref{fig:num_rgb}(f).
The second numerical result corresponded to DoF-NeRF, in which the captured images, $\bm{C}_\text{cap}$, generated using the lens camera model were used to jointly train the network and the optical parameters under the lens camera model.
After the two-stage training described in Subsection~\ref{subsec:two_stage}, the blur convolution process in Eq.~\eqref{eq:lens_rendering} was removed, and the color radiance~$\bm{C}_\text{pin}$ was computed as the EDoF novel-view image.
In this case, the defocus blur was alleviated compared with the first case, but it still remained in the novel-view image, as shown in Fig.~\ref{fig:num_rgb}(g).
The third numerical result corresponded to EDoF-NeRF, in which the captured images, $\bm{C}_\text{cap}$, generated using the coded aperture camera model were used to jointly train the network and the optical parameters under the coded aperture camera model.
Following the same procedure as in the second case, the color radiance~$\bm{C}_\text{pin}$ was computed as the EDoF novel-view image.
In this case, the defocus blur was sufficiently compensated compared with the first and second cases, as shown in Fig.~\ref{fig:num_rgb}(h).
Visualization~1 summarizes these results in video format.

The reconstruction performance of the three NeRF variants was quantitatively evaluated using the peak signal-to-noise ratio~(PSNR) and the structural similarity index~(SSIM) between the reconstructed novel-view images and the reference novel-view images~\cite{wang2004Image}.
The resulting metrics are summarized in Table~\ref{tab:num_rgb}, where EDoF-NeRF, corresponding to Fig.~\ref{fig:num_rgb}(h), outperforms both NeRF, corresponding to Fig.~\ref{fig:num_rgb}(f), and DoF-NeRF, corresponding to Fig.~\ref{fig:num_rgb}(g).
These qualitative and quantitative results demonstrate that incorporating a coded aperture into the NeRF pipeline effectively mitigates the irreversible information loss inherent in conventional apertures---particularly at high spatial frequencies---thereby enabling high-fidelity NeRF reconstruction from defocused inputs.

\begin{table}[tbp]
  \centering
  \caption{Quantitative evaluation of the reconstructed novel-view images in the numerical demonstration.}
  \label{tab:num_rgb}
  \begin{tabular}{c | c c}
    \hline
    Evaluated model & PSNR~[dB]$\uparrow$ & SSIM$\uparrow$ \\
    \hline
    NeRF~(Fig.~\ref{fig:num_rgb}(f)) & 15.6 & 0.324 \\
    DoF-NeRF~(Fig.~\ref{fig:num_rgb}(g)) & 19.0 & 0.551 \\
    EDoF-NeRF~(Fig.~\ref{fig:num_rgb}(h)) & \textbf{20.4} & \textbf{0.636} \\
    \hline
  \end{tabular}
\end{table}

\begin{figure}[tbp]
  \centering
  \begin{subfigure}[t]{0.34\linewidth}
    \centering
    \includegraphics[height=4.5cm]{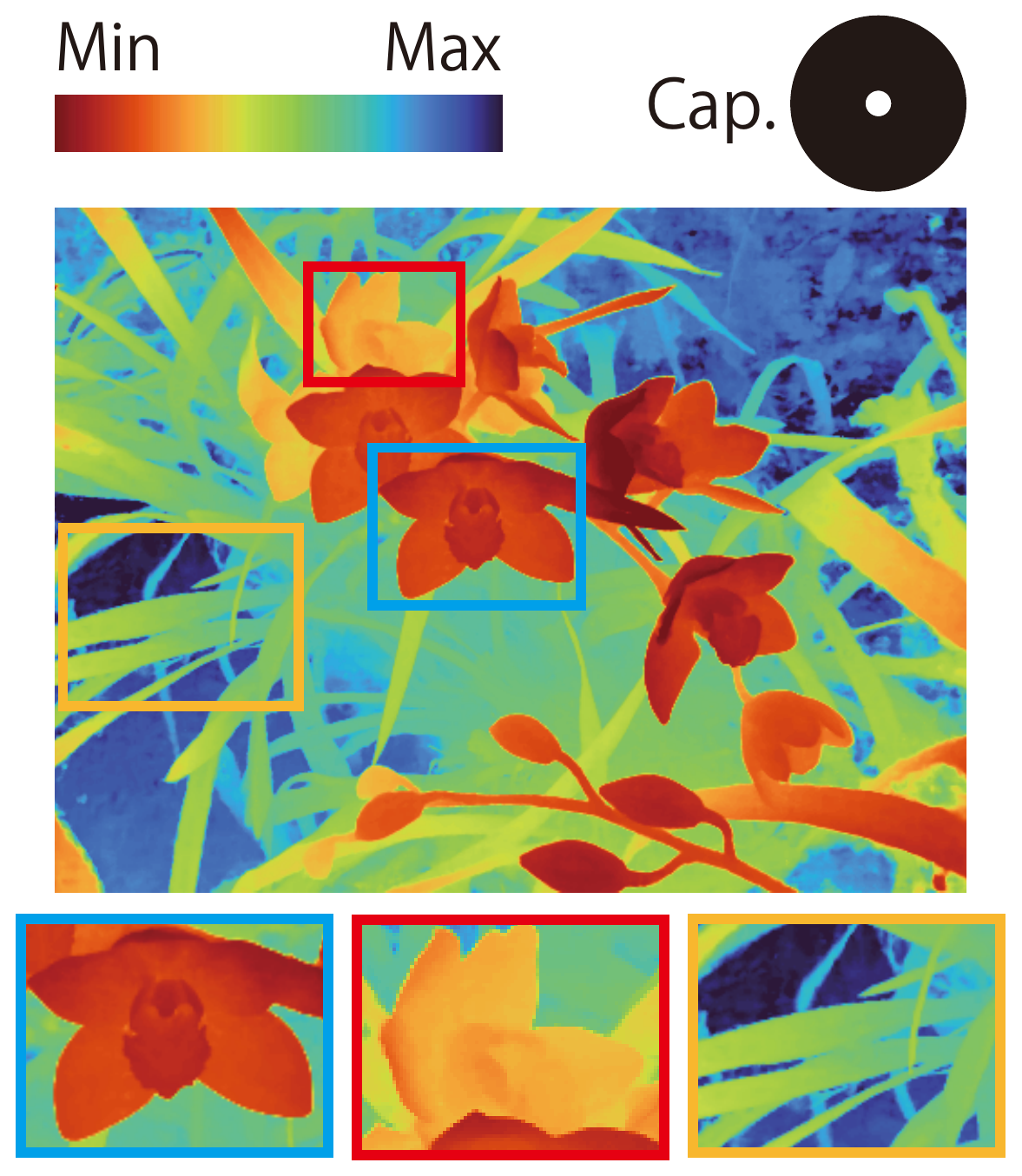}
    \caption*{(a)}
  \end{subfigure}
  \begin{subfigure}[t]{0.34\linewidth}
    \centering
    \includegraphics[height=4.5cm]{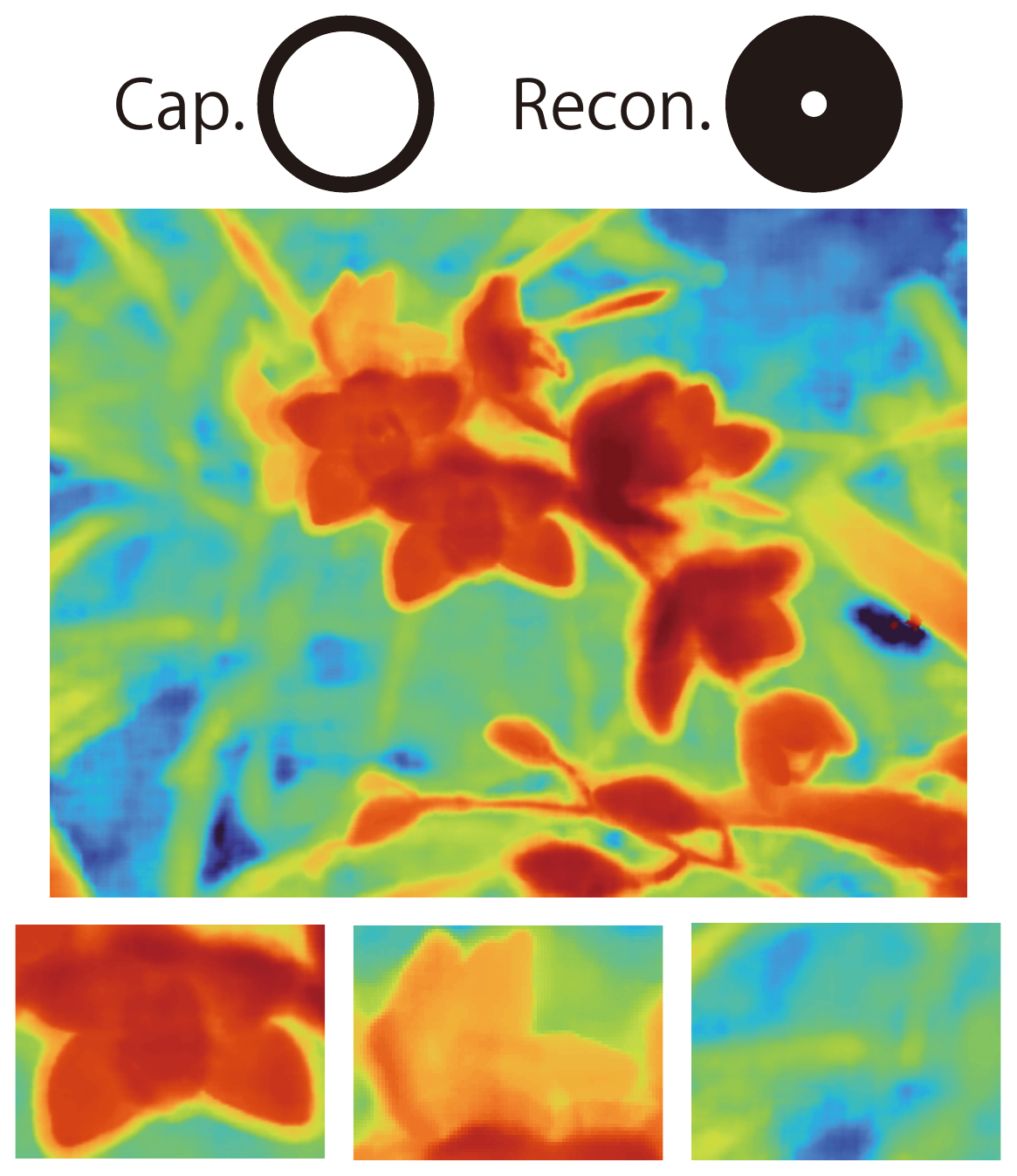}
    \caption*{(b)}
  \end{subfigure}

  \vspace{2mm}
  
  \begin{subfigure}[t]{0.34\linewidth}
    \centering
    \includegraphics[height=4.5cm]{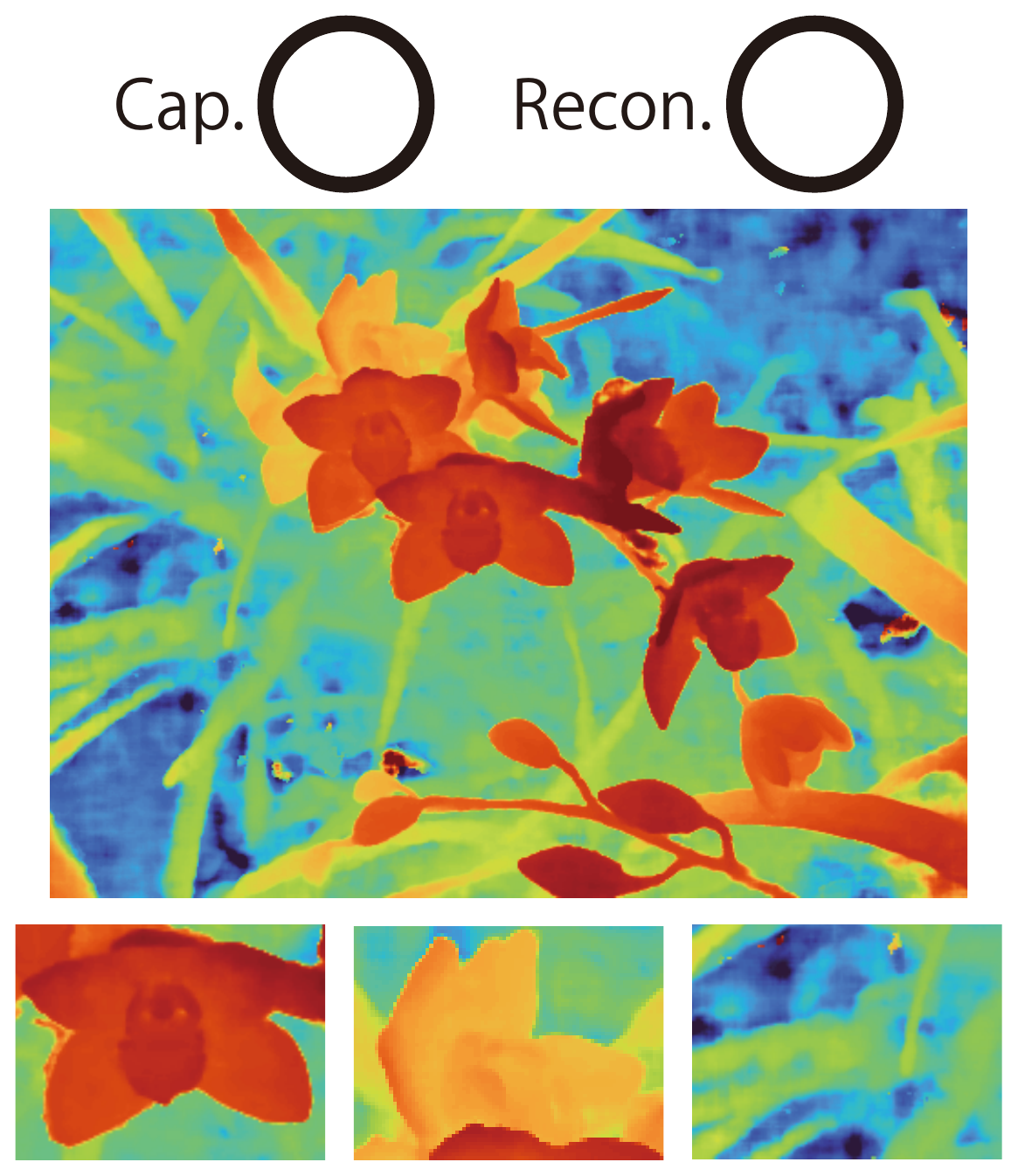}
    \caption*{(c)}
  \end{subfigure}
  \begin{subfigure}[t]{0.34\linewidth}
    \centering
    \includegraphics[height=4.5cm]{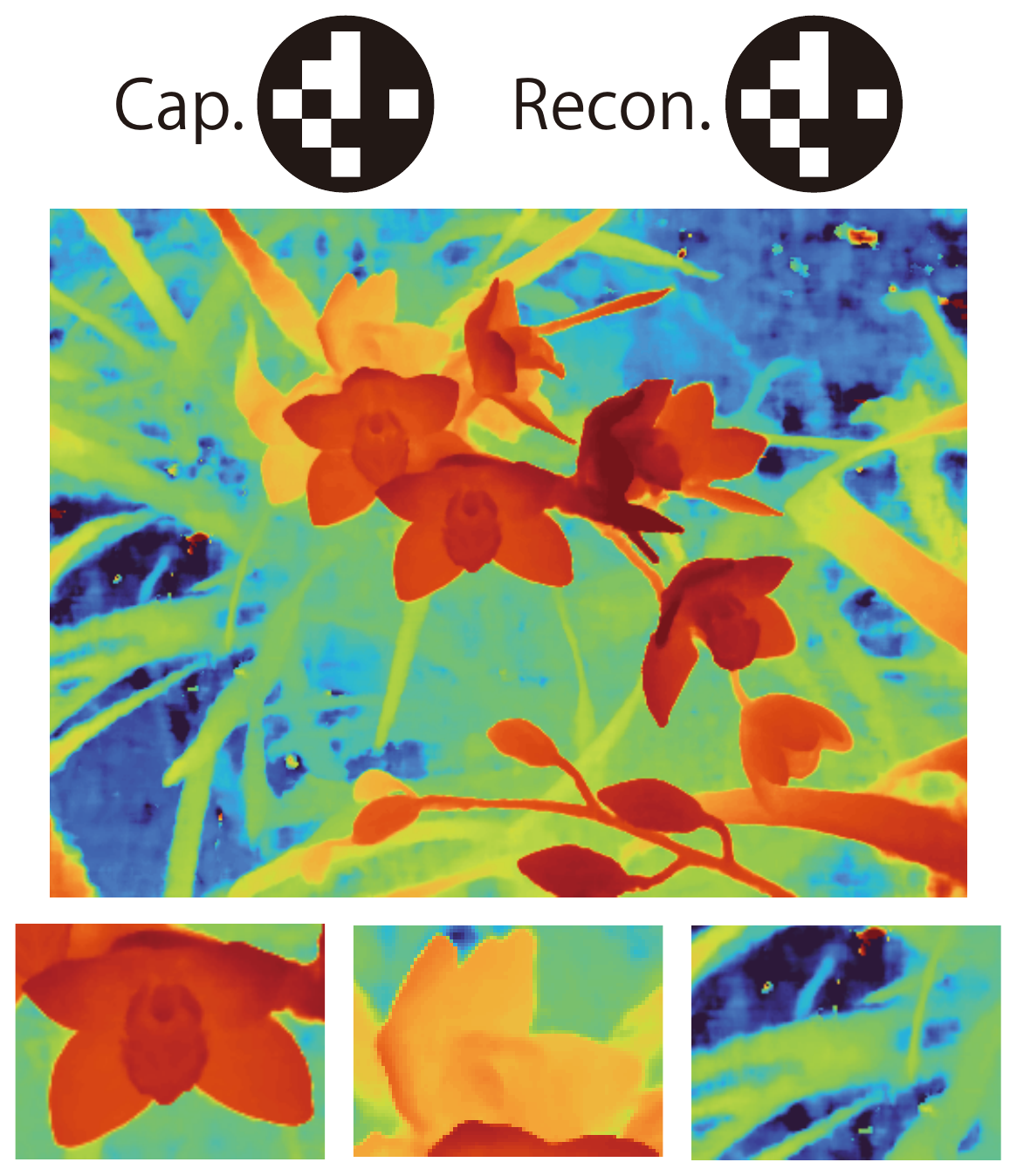}
    \caption*{(d)}
  \end{subfigure}
  \caption{Depth maps in the numerical demonstration.
(a)~Reference depth map rendered from the images generated by the pinhole camera model in Fig.~\ref{fig:num_rgb}(a).
Rendered depth maps using the (b)~pinhole and (c)~lens camera models from the images generated by the lens camera model in Figs.~\ref{fig:num_rgb}(b) and \ref{fig:num_rgb}(d).
(d)~Rendered depth map using the coded aperture camera model from the images generated by the coded aperture camera model in Figs.~\ref{fig:num_rgb}(c) and \ref{fig:num_rgb}(e).
The color bar for the depth maps is shown in the upper left of (a).
The zoom-in views, pupil illustrations, and abbreviations follow those in Fig.~\ref{fig:num_rgb}.
  }
  \label{fig:num_depth}
\end{figure}

Furthermore, we evaluated the geometric reconstruction performance by rendering depth maps based on the proxy depth~$R_\text{proxy}(\bm{p})$ defined in Eq.~\eqref{eq:proxy_depth}.
The depth map produced by the reference NeRF is shown in Fig.~\ref{fig:num_depth}(a).
The depth maps estimated by NeRF, DoF-NeRF, and EDoF-NeRF are presented in Figs.~\ref{fig:num_depth}(b)--\ref{fig:num_depth}(d), respectively.
Among the three methods, EDoF-NeRF produces the sharpest depth map with the fewest reconstruction artifacts.  
The PSNR and SSIM values of the depth maps are summarized in Table~\ref{tab:num_depth}, where these metrics were calculated after linearly normalizing the depth maps using their global minimum and maximum values.
These results confirm that EDoF-NeRF possesses superior capability for precise scene geometry estimation, even when the input images are captured under shallow-DoF conditions.

\begin{table}[tbp]
  \centering
  \caption{Quantitative evaluation of the estimated depth maps in the numerical demonstration.}
  \label{tab:num_depth}
  \begin{tabular}{c | c c}
    \hline
    Evaluated model & PSNR~[dB]$\uparrow$ & SSIM$\uparrow$ \\
    \hline
    NeRF~(Fig.~\ref{fig:num_depth}(b))      & 17.8 & 0.563 \\
    DoF-NeRF~(Fig.~\ref{fig:num_depth}(c))  & 21.7 & 0.722 \\
    EDoF-NeRF~(Fig.~\ref{fig:num_depth}(d)) & \textbf{23.1} & \textbf{0.765} \\
    \hline
  \end{tabular}
\end{table}

\subsection{Optical demonstration}
\label{subsec:opt_demo}

\begin{figure}[tbp]
  \centering
  \begin{subfigure}[t]{1.0\linewidth}
    \centering
    \includegraphics[width=0.34\linewidth]{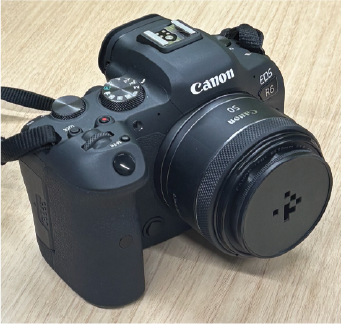}
    \caption{} 
  \end{subfigure}
  
  \vspace{1mm} 
  
  \begin{subfigure}[t]{0.34\linewidth}
    \centering
    \includegraphics[width=\linewidth]{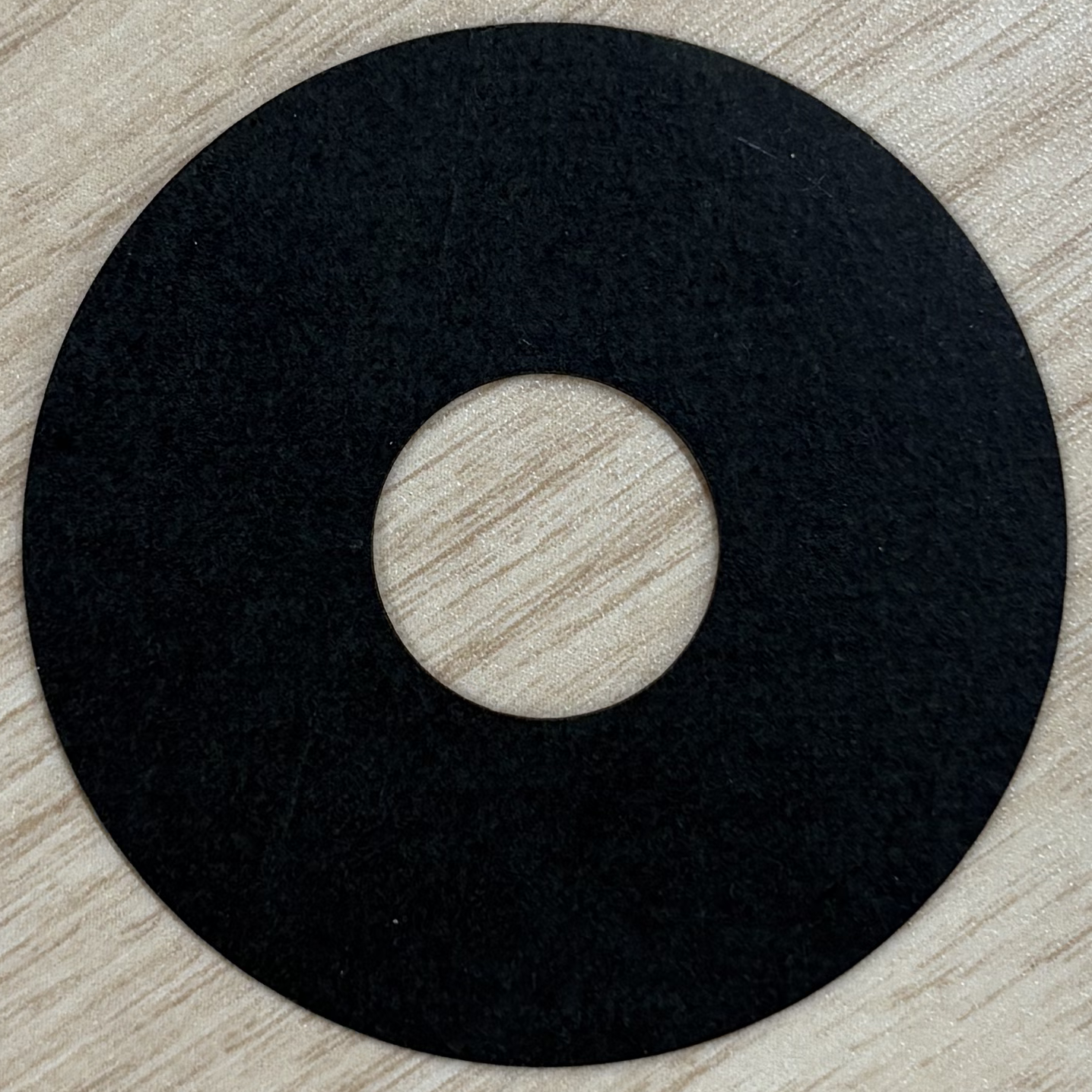}
    \caption{} 
  \end{subfigure}
  \hspace{10mm} 
  \begin{subfigure}[t]{0.34\linewidth}
    \centering
    \includegraphics[width=\linewidth]{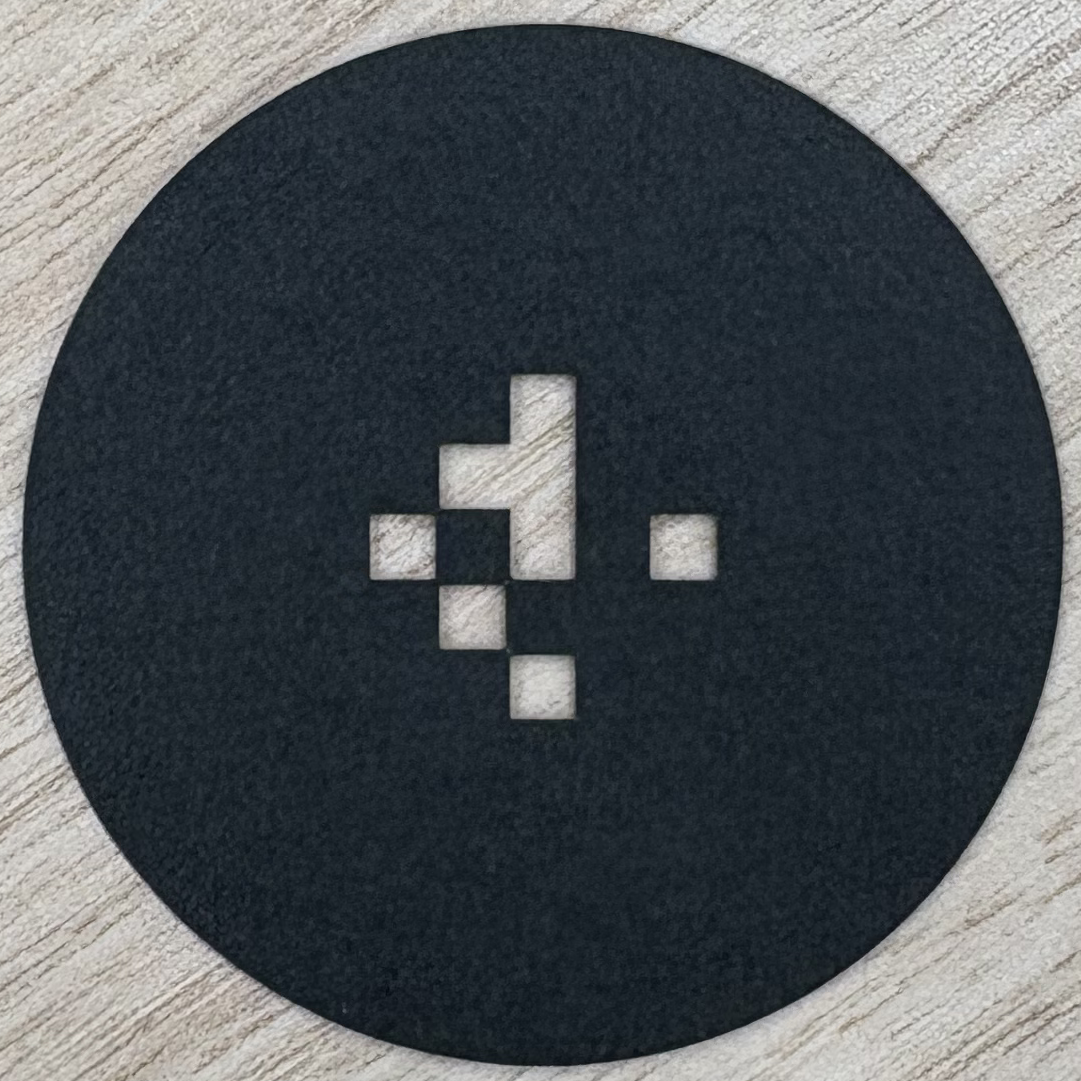}
    \caption{} 
  \end{subfigure}

  \caption{Experimental camera setup for optical validation.
(a) Camera equipped with the coded aperture.
(b) Circular aperture.
(c) Coded aperture.}
  \label{fig:setup}
\end{figure}

We further validated the effectiveness of EDoF-NeRF in a real-world environment using a commercial camera~(Canon EOS R6 with a Canon RF50mm F1.8 STM prime lens) equipped with external masks.  
The external masks with 0.5-mm-thick black cardstock were fabricated using a laser cutter and attached to the front of the lens using an aperture filter~(GIZMON Bokeh Freedom Filter), as shown in Fig.~\ref{fig:setup}, to implement the circular aperture in Fig.~\ref{fig:model_lens} and the coded aperture in Fig.~\ref{fig:model_coded}.  
The exposure time in each case was set using the camera's auto-exposure mode.  
To obtain reference all-in-focus images under an approximate pinhole camera condition corresponding to the pinhole camera in Fig.~\ref{fig:model_pinhole} for quantitative evaluation, the pupil aperture inside the prime lens was set to its minimum~(F22) with the circular aperture, along with a long exposure time.

\begin{figure}[tbp]
  \centering
  
  \begin{subfigure}[t]{0.32\linewidth}
    \centering
    \rule{0pt}{4.5cm} 
    \vspace{-2mm}
  \end{subfigure}
  \hfill
  \begin{subfigure}[t]{0.32\linewidth}
    \centering
    \includegraphics[height=4.5cm]{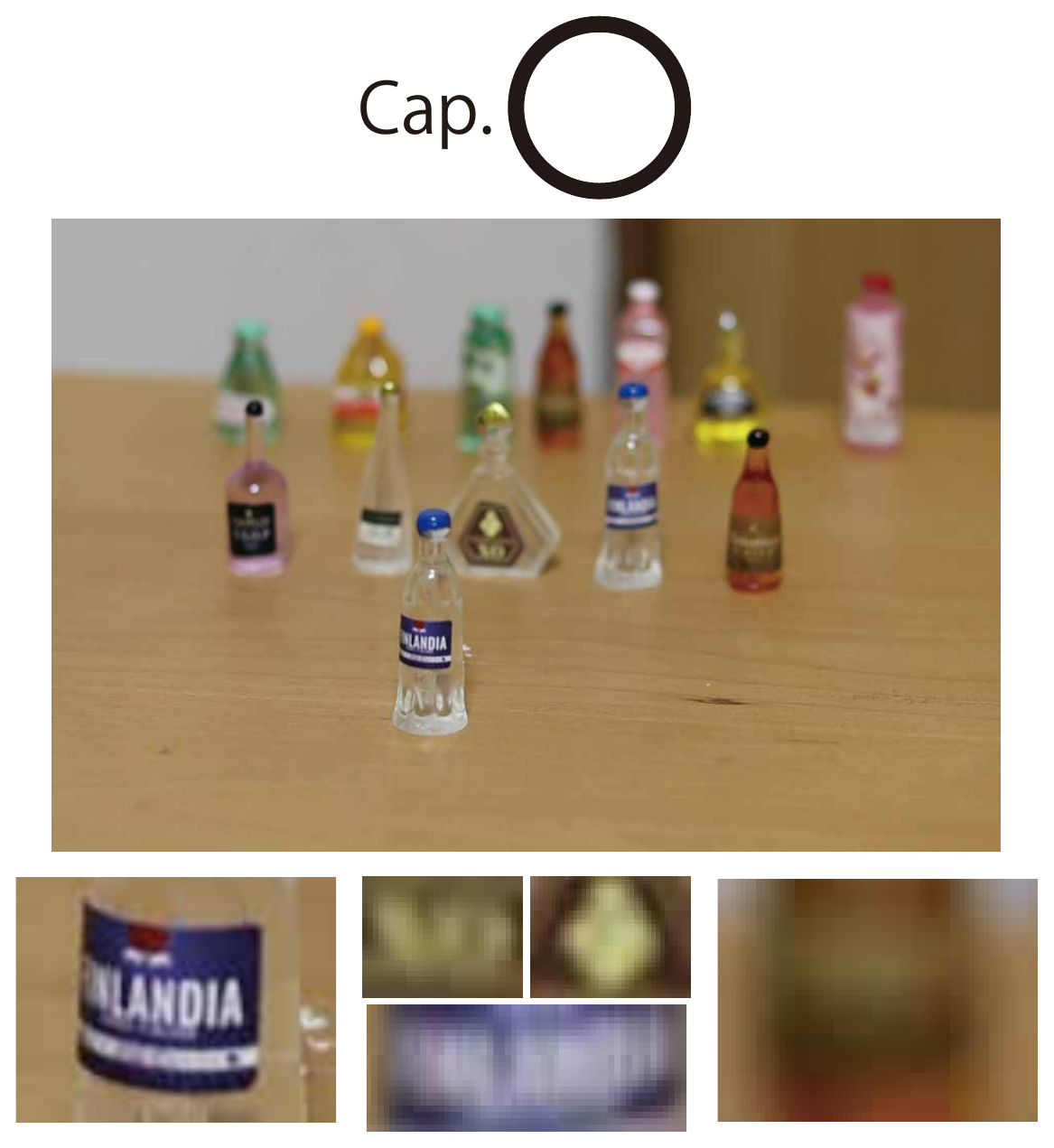}
    \vspace{-2mm}
    \caption*{(b)}
  \end{subfigure}
  \hfill
  \begin{subfigure}[t]{0.32\linewidth}
    \centering
    \includegraphics[height=4.5cm]{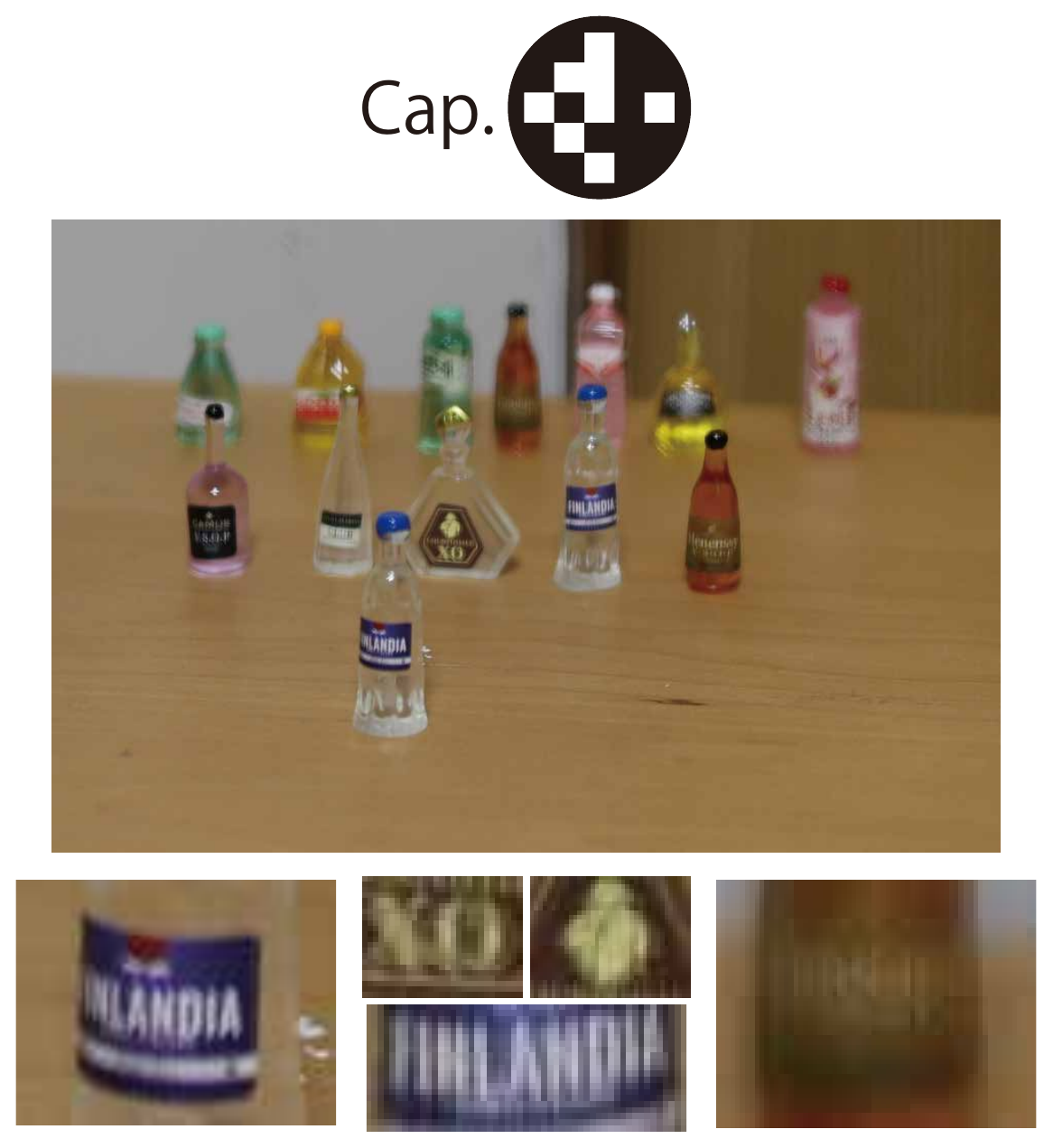}
    \vspace{-2mm}
    \caption*{(c)}
  \end{subfigure}

  \vspace{4mm}

  \begin{subfigure}[t]{0.32\linewidth}
    \centering
    \includegraphics[height=4.5cm]{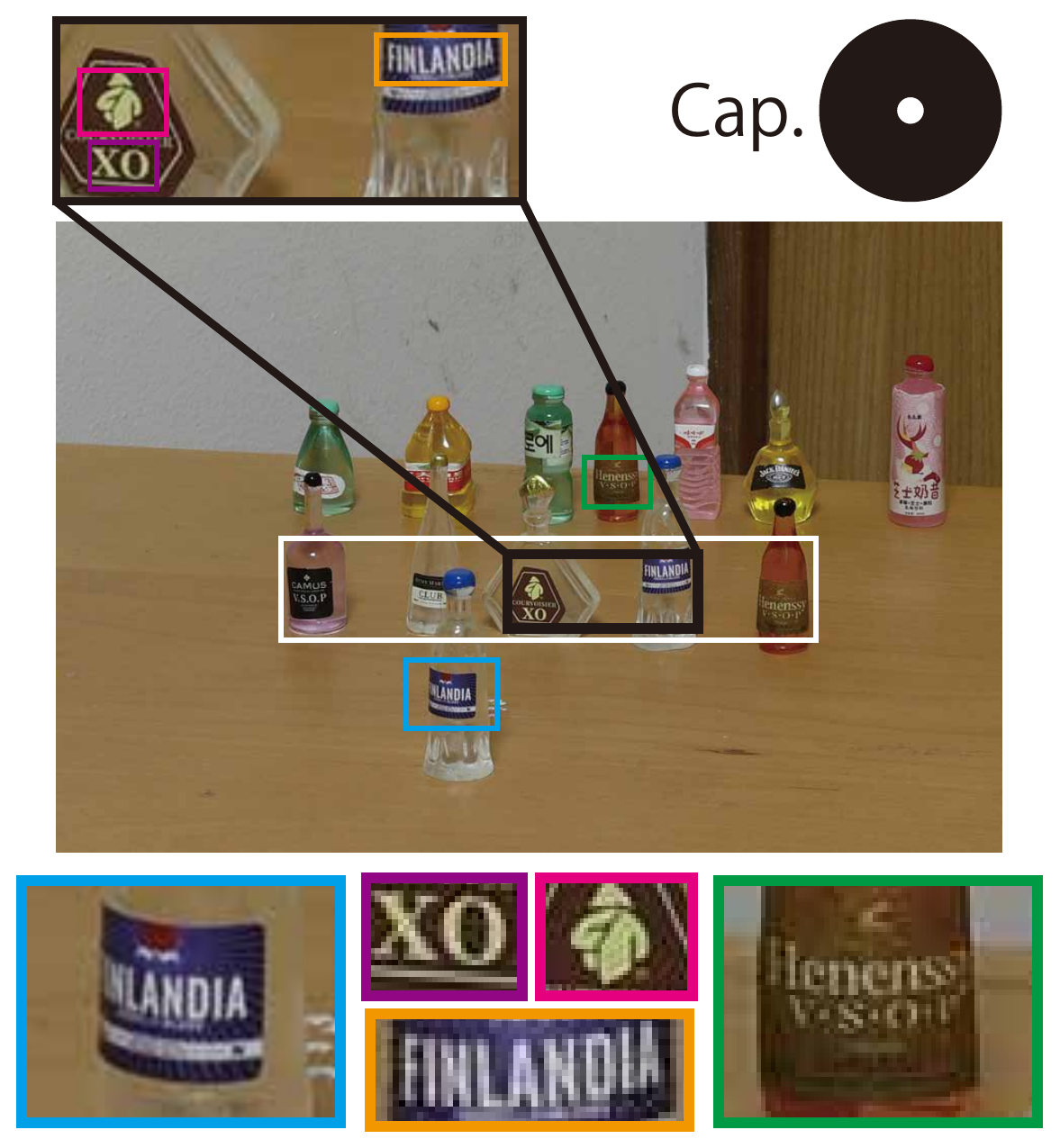}
    \vspace{-2mm}
    \caption*{(a)}
  \end{subfigure}
  \hfill
  \begin{subfigure}[t]{0.32\linewidth}
    \centering
    \includegraphics[height=4.5cm]{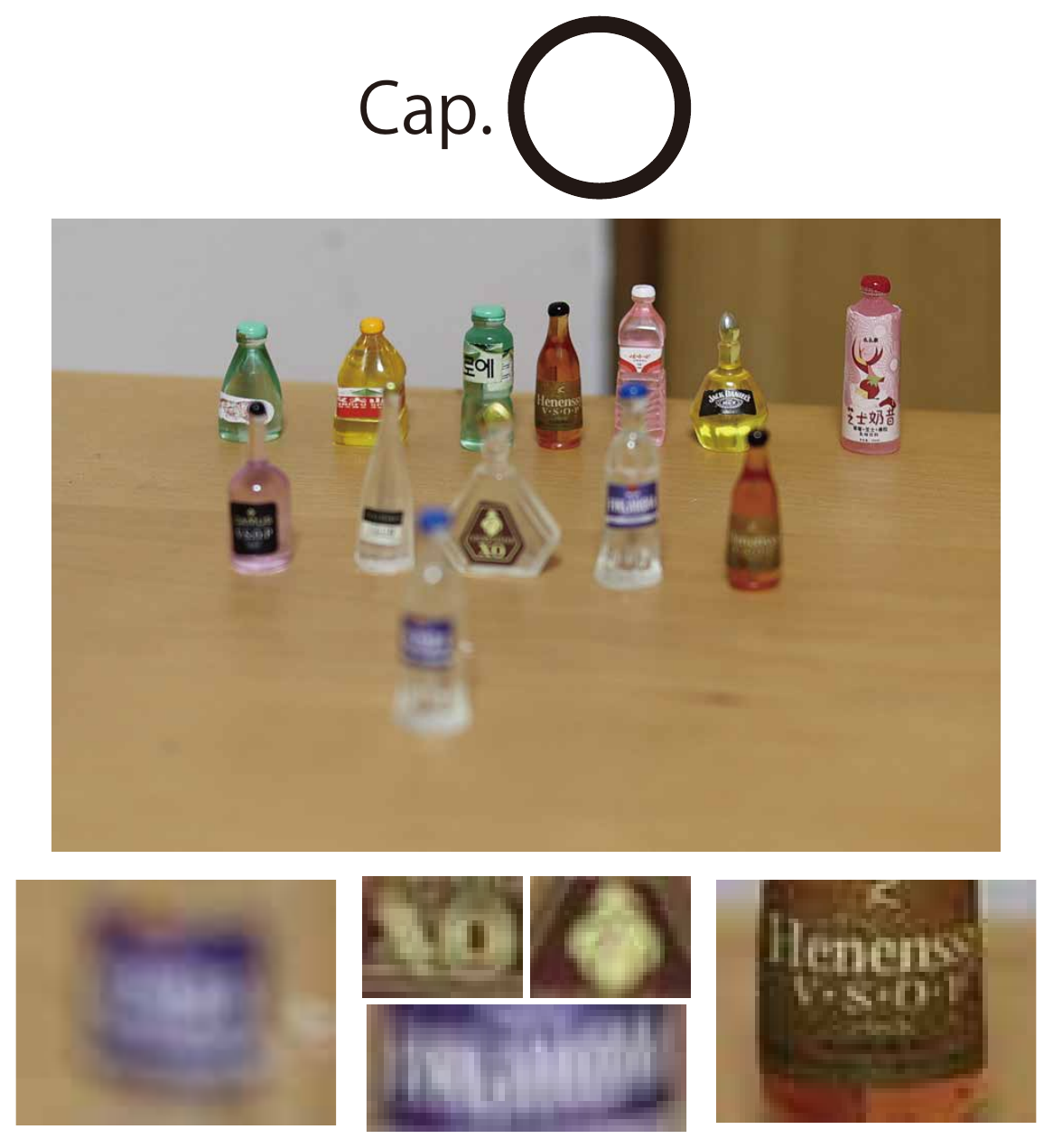}
    \vspace{-2mm}
    \caption*{(d)}
  \end{subfigure}
  \hfill
  \begin{subfigure}[t]{0.32\linewidth}
    \centering
    \includegraphics[height=4.5cm]{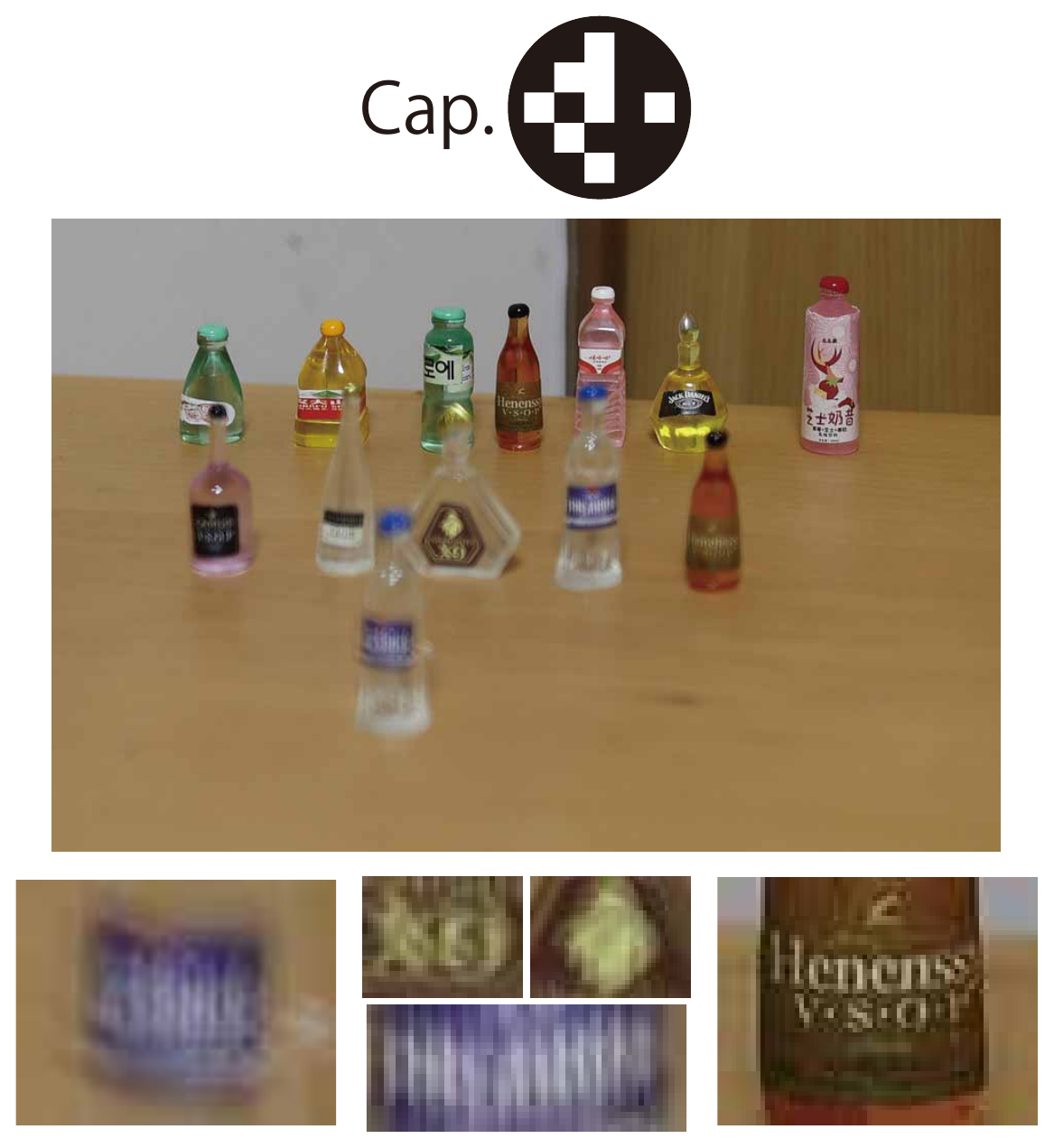}
    \vspace{-2mm}
    \caption*{(e)}
  \end{subfigure}

  \vspace{6mm}

  \begin{subfigure}[t]{0.32\linewidth}
    \centering
    \includegraphics[height=4.5cm]{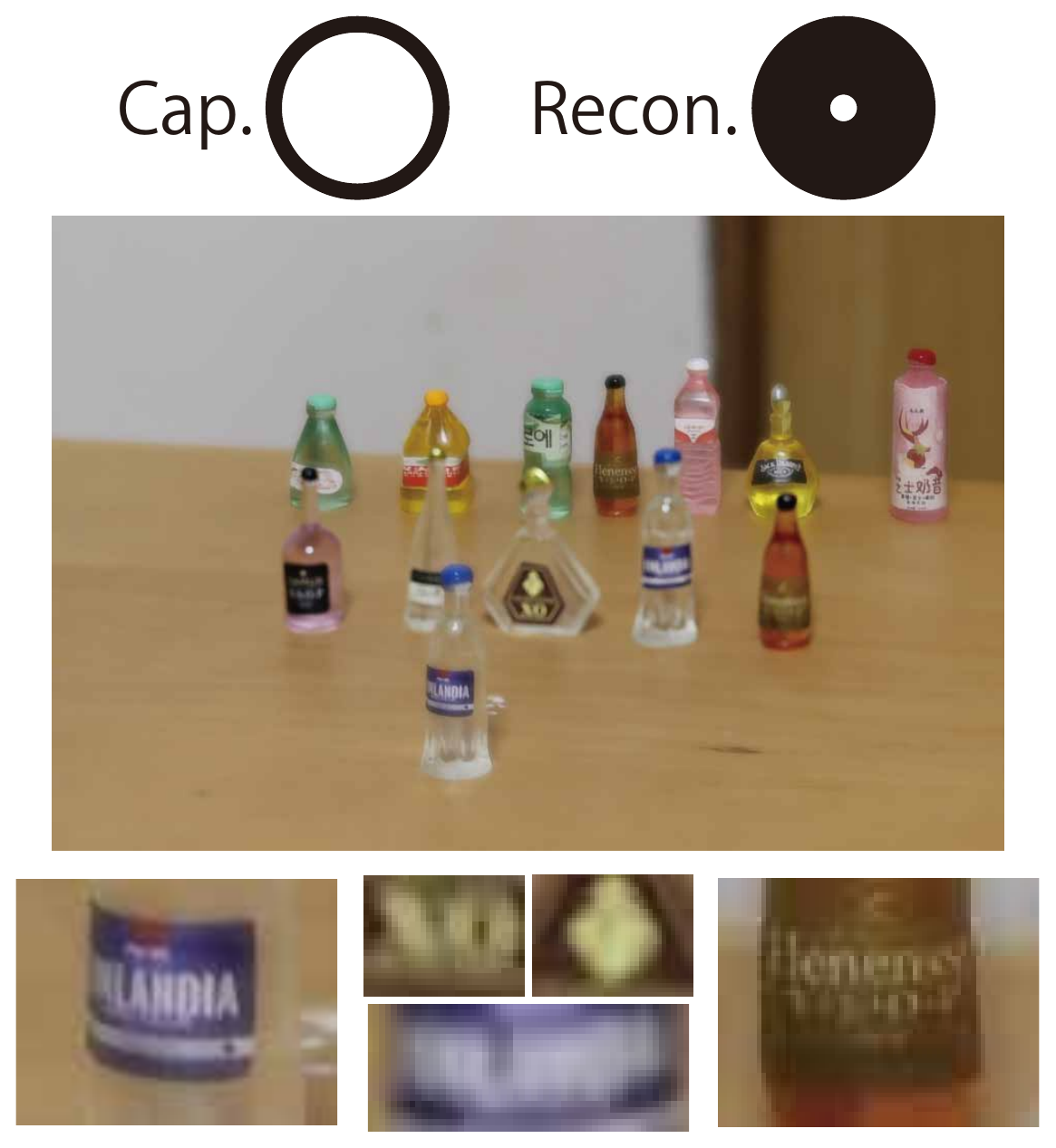}
    \vspace{-2mm}
    \caption*{(f)}
  \end{subfigure}
  \hfill
  \begin{subfigure}[t]{0.32\linewidth}
    \centering
    \includegraphics[height=4.5cm]{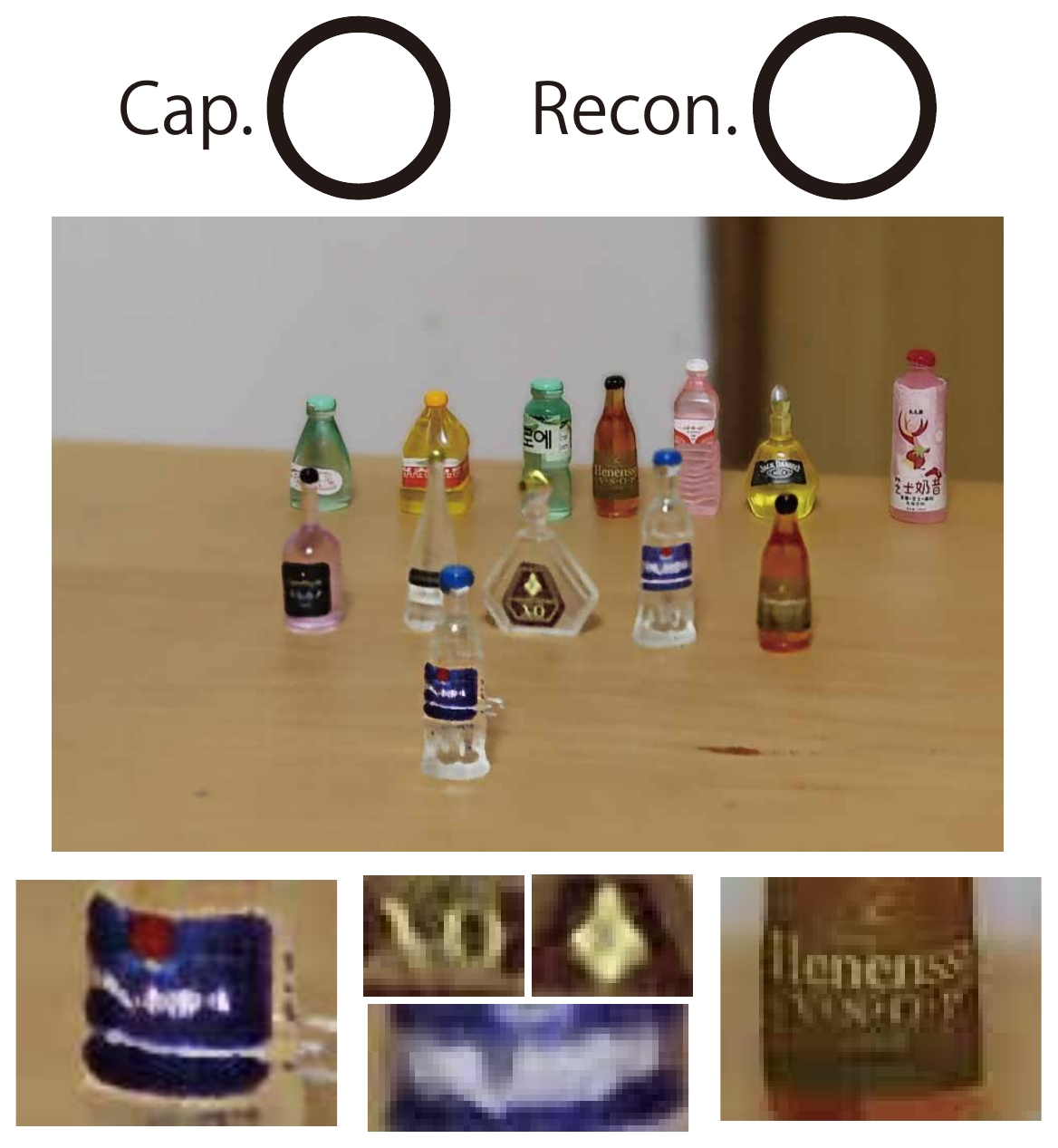}
    \vspace{-2mm}
    \caption*{(g)}
  \end{subfigure}
  \hfill
  \begin{subfigure}[t]{0.32\linewidth}
    \centering
    \includegraphics[height=4.5cm]{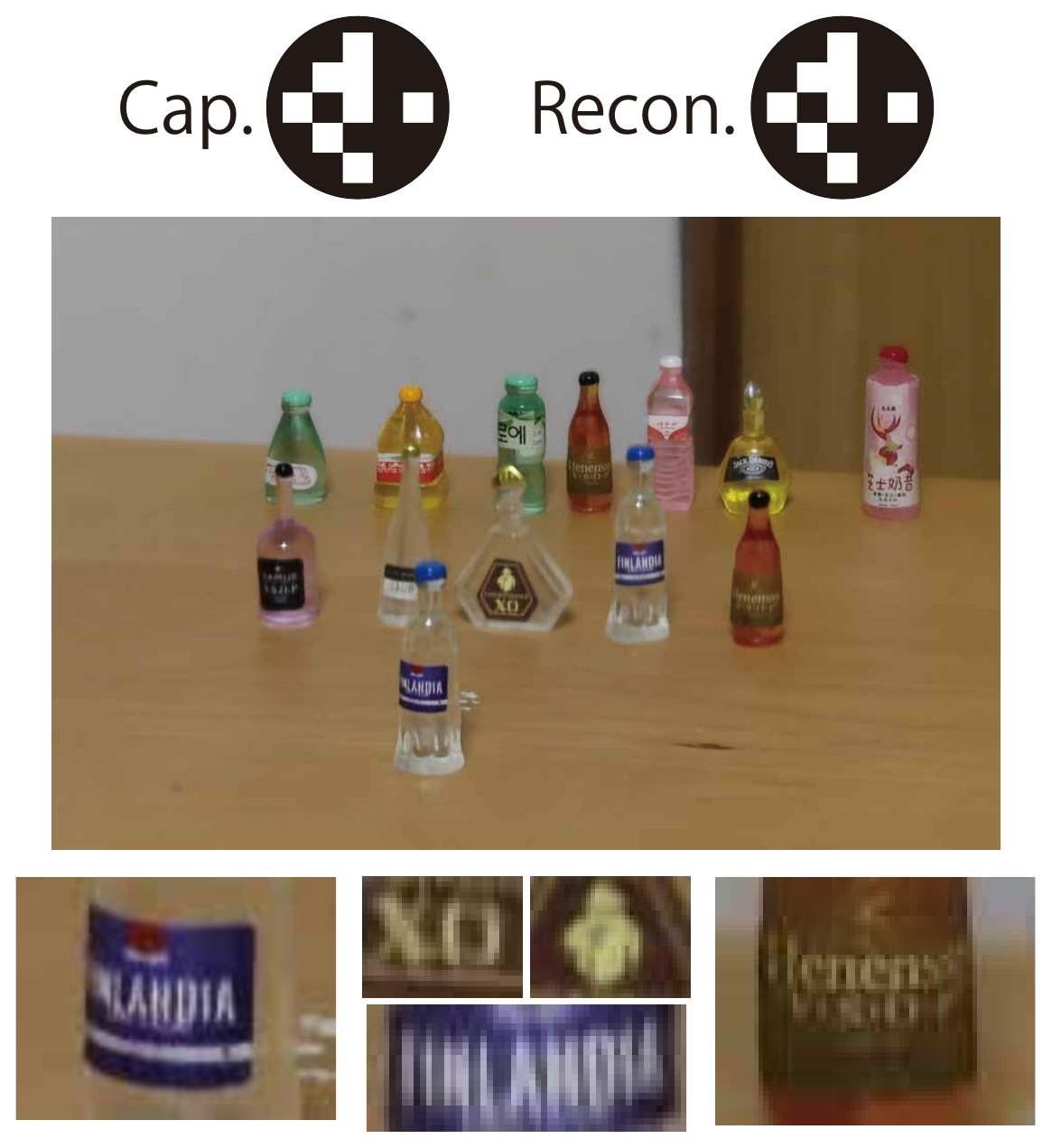}
    \vspace{-2mm}
    \caption*{(h)}
  \end{subfigure}

  \vspace{2mm}
  \caption{
  Optical demonstration results.
  (a)~Reference all-in-focus image captured under the approximate pinhole camera condition by stopping down the aperture to its minimum size.
  Foreground-focused captured images under the (b)~lens and (c)~coded aperture camera conditions.
  Background-focused captured images under the (d)~lens and (e)~coded aperture camera conditions.
  Reconstruction results from (d) using the (f)~pinhole and (g)~lens camera models.
  (h)~Reconstruction result from (e) using the coded aperture camera model.
  The zoom-in views, pupil illustrations, and abbreviations follow those in Fig.~\ref{fig:num_rgb}.
  }
  \label{fig:opt_rgb}
\end{figure}

In the optical demonstration, the target scene consisted of miniature bottles placed at different depths on a desk, as shown in Fig.~\ref{fig:opt_rgb}(a), which serves as the reference all-in-focus image obtained using the pinhole camera. 
For each of the circular and coded apertures, we captured 24~images, corresponding to 12~viewpoints and two focus conditions~(foreground and background), as shown in Figs.~\ref{fig:opt_rgb}(b) and (d), and Figs.~\ref{fig:opt_rgb}(c) and (e), respectively.
Before training, we used COLMAP with the images captured under each aperture to estimate the camera poses~(positions and orientations), following the original NeRF pipeline~\cite{schonberger2016Structure}.  
The reconstructed novel-view image using NeRF is shown in Fig.~\ref{fig:opt_rgb}(f), where the images captured with the circular aperture were used to optimize only the network parameters under the pinhole camera model.  
In this case, the defocus on the mid-ground object, which was out of focus during image capture, remains.  
The reconstructed novel-view image using DoF-NeRF is shown in Fig.~\ref{fig:opt_rgb}(g), where the images captured with the circular aperture were used to jointly optimize the network and the optical parameters under the lens camera model.  
In this case, the defocus on the mid-ground object is alleviated but still insufficient, and artifacts appear on the foreground object.  
The reconstructed novel-view image using EDoF-NeRF is shown in Fig.~\ref{fig:opt_rgb}(h), where the images captured with the coded aperture were used to jointly optimize the network and the optical parameters under the coded aperture camera model.  
In this case, the defocus on the mid-ground object is effectively compensated, and the reconstruction remains stable across all depths.
Visualization~2 summarizes these results in video format.

The quantitative evaluation of the novel-view images reconstructed in the optical demonstration is summarized in Table~\ref{tab:opt_rgb}.
The evaluation was conducted on the region enclosed by the white bounding box in Fig.~\ref{fig:opt_rgb}(a).
The bottles located within this region were defocused in all training images captured under each aperture condition.
Therefore, this region is suitable for evaluating the deblurring performance of each method.
To compensate for positional shifts and brightness variations arising from different aperture conditions, affine transformations and RGB scaling coefficients were applied to the novel-view images reconstructed by NeRF, DoF-NeRF, and EDoF-NeRF, such that the white-bounding-box region in each reconstructed image was aligned with that in the reference image.
These alignment parameters were estimated using the Adam optimizer to minimize the MSE between the reconstructed novel-view image and the reference image over the white-bounding-box region.
The PSNR and SSIM results further support the advantage of EDoF-NeRF over NeRF and DoF-NeRF.

\begin{figure}[tbp]
  \centering
  \begin{subfigure}[t]{0.34\linewidth}
    \centering
    \includegraphics[height=4.5cm]{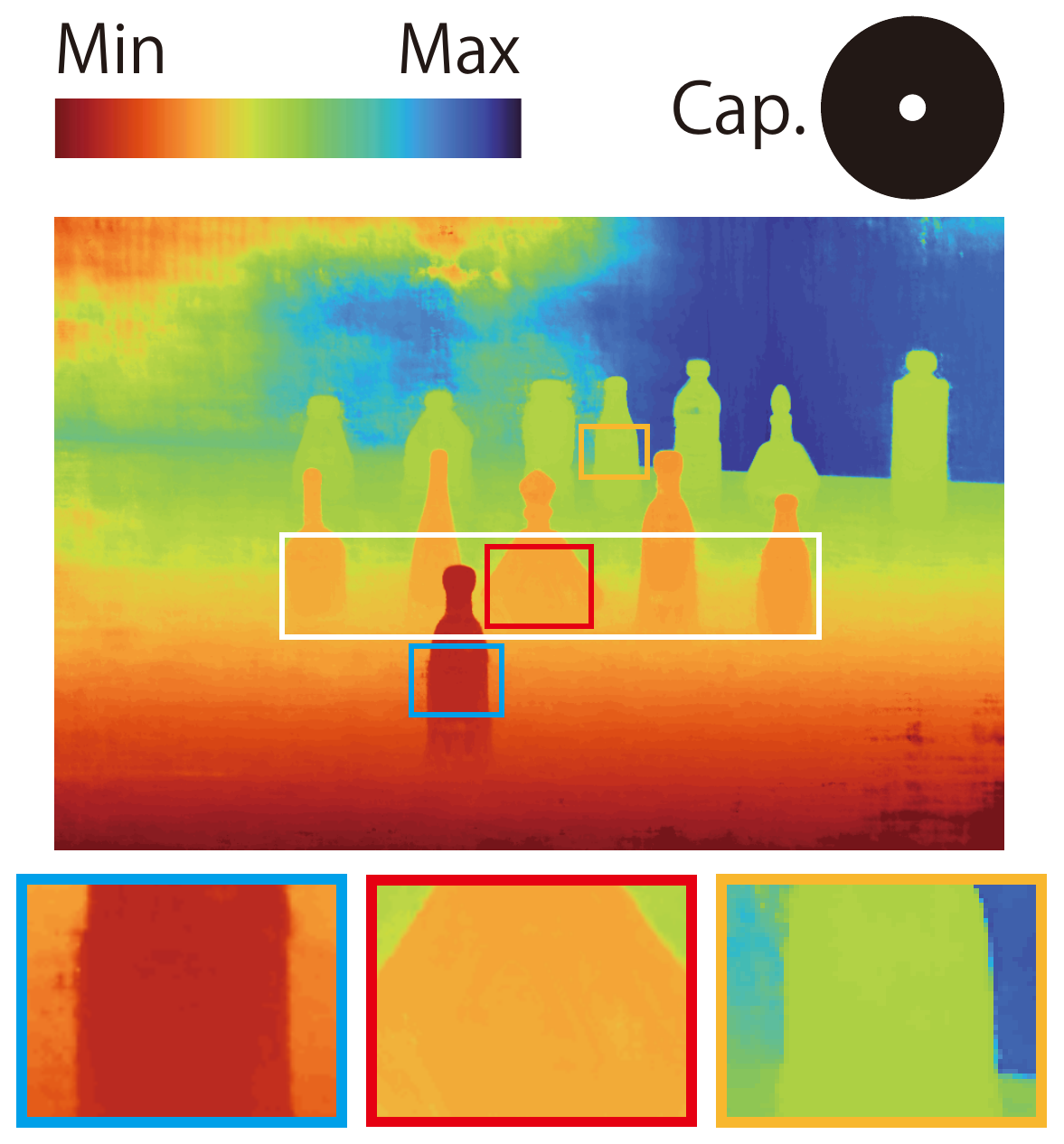}
    \caption*{(a)}
  \end{subfigure}
  \begin{subfigure}[t]{0.34\linewidth}
    \centering
    \includegraphics[height=4.5cm]{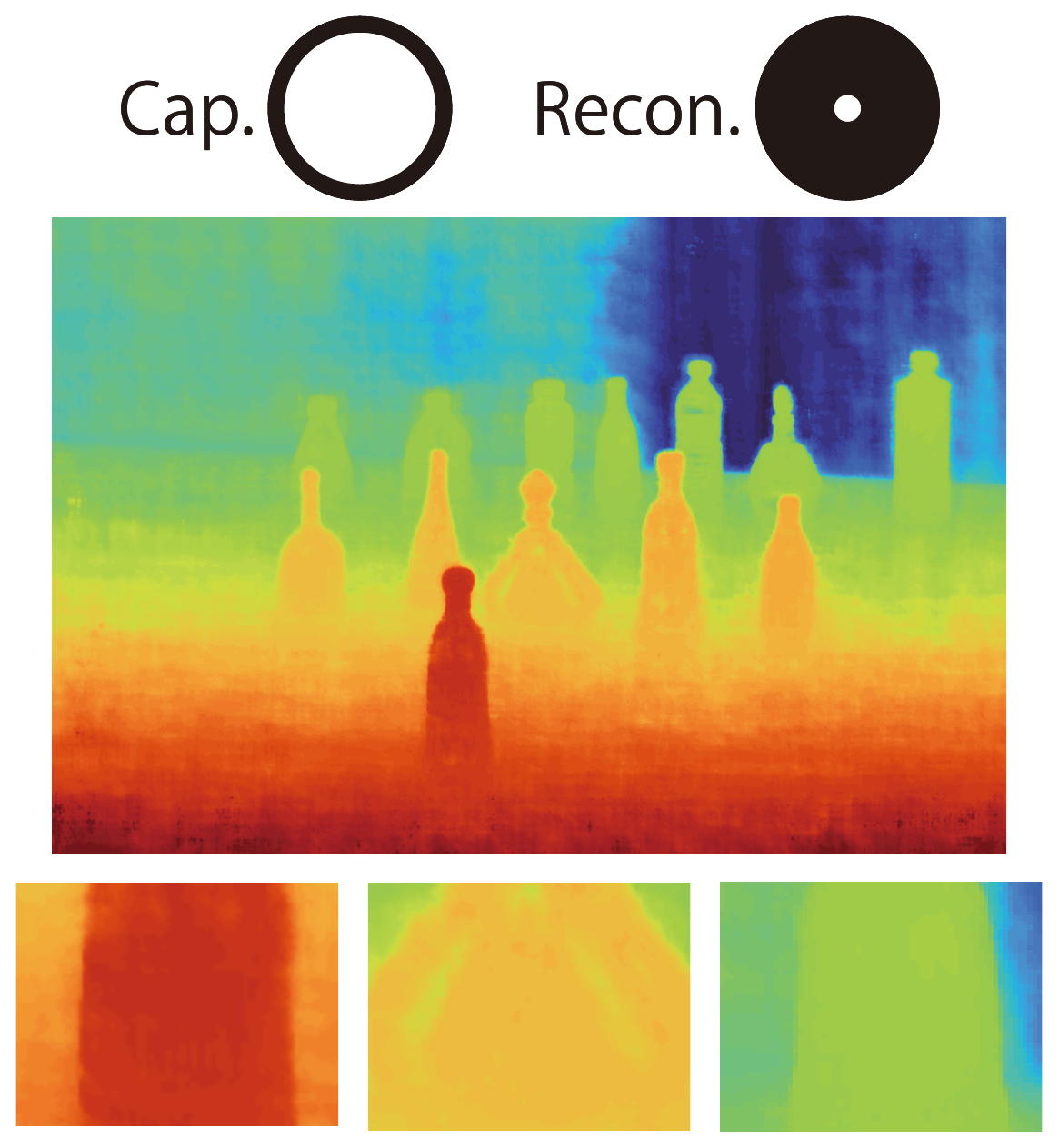}
    \caption*{(b)}
  \end{subfigure}

  \vspace{2mm}
  
  \begin{subfigure}[t]{0.34\linewidth}
    \centering
    \includegraphics[height=4.5cm]{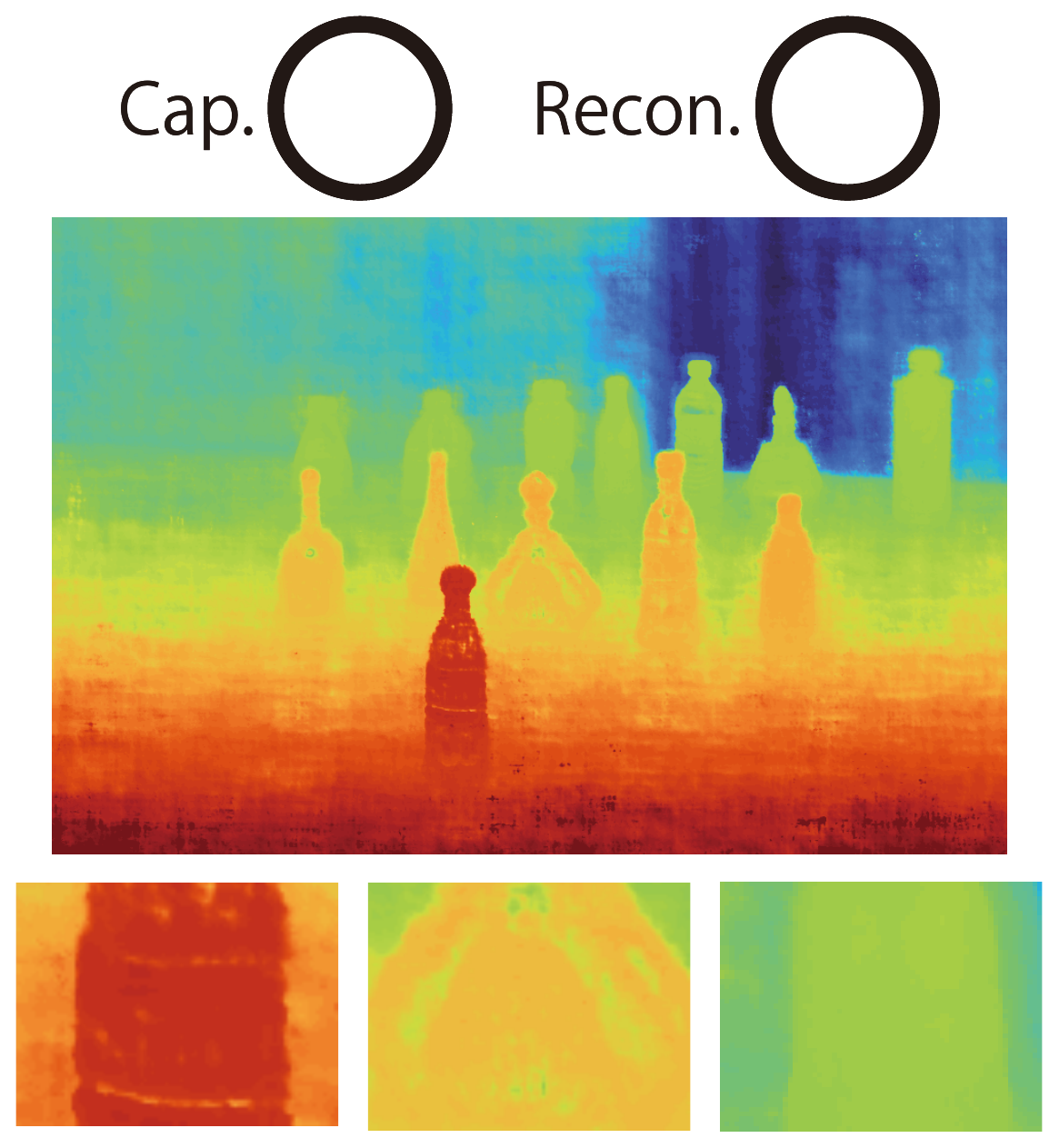}
    \caption*{(c)}
  \end{subfigure}
  \begin{subfigure}[t]{0.34\linewidth}
    \centering
    \includegraphics[height=4.5cm]{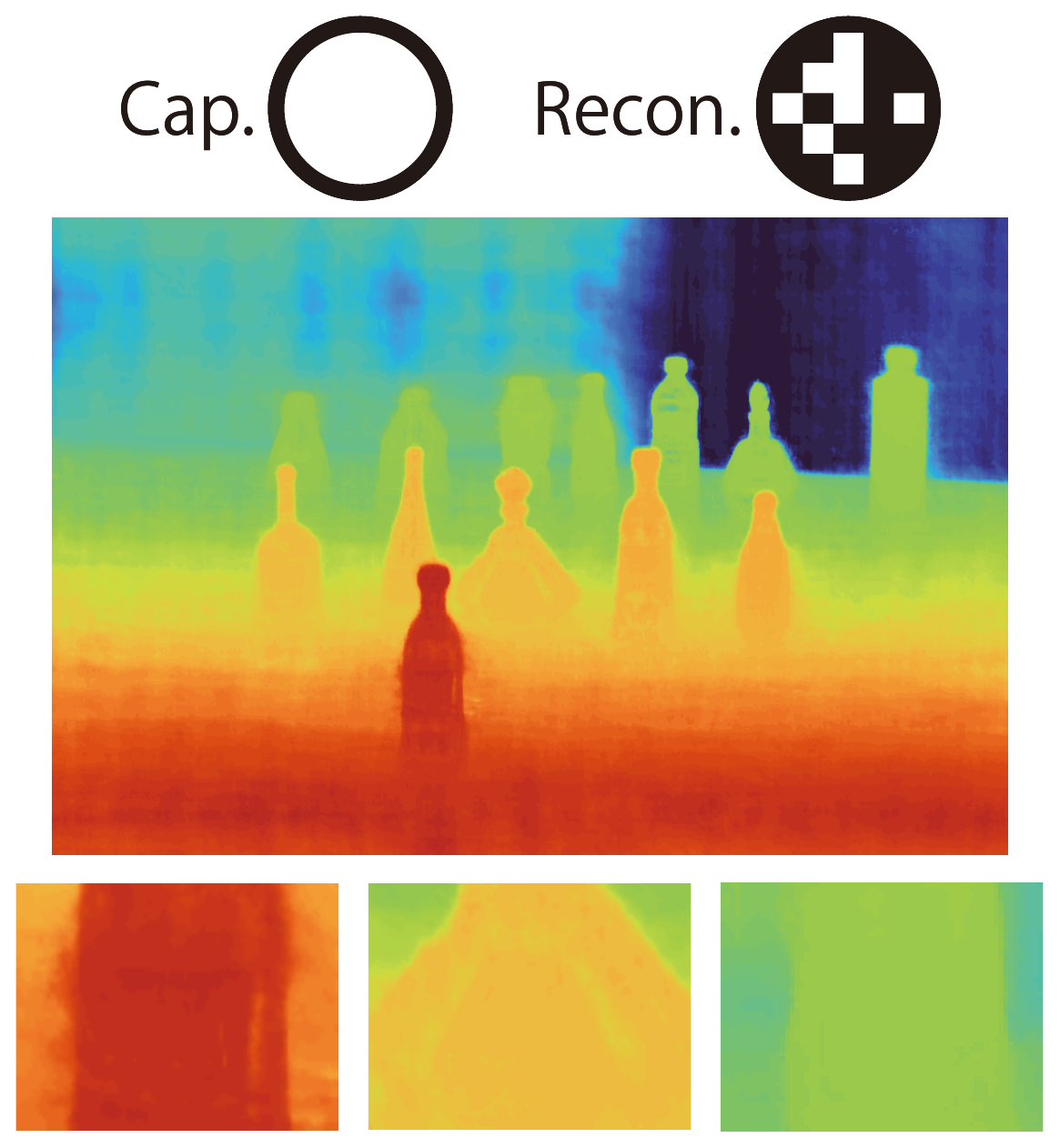}
    \caption*{(d)}
  \end{subfigure}
  \caption{Depth maps in the optical demonstration.
(a)~Reference depth map rendered from the images captured under the approximate pinhole camera condition in Fig.~\ref{fig:opt_rgb}(a).
Rendered depth maps using the (b)~pinhole and (c)~lens camera models from the images captured under the lens camera condition in Figs.~\ref{fig:opt_rgb}(b) and \ref{fig:opt_rgb}(d).
(d)~Rendered depth map using the coded aperture camera model from the images captured under the coded aperture camera condition in Figs.~\ref{fig:opt_rgb}(c) and \ref{fig:opt_rgb}(e).
The color bar for the depth maps is shown in the upper left of (a).
The zoom-in views, pupil illustrations, and abbreviations follow those in Fig.~\ref{fig:num_rgb}.
  }
  \label{fig:opt_depth}
\end{figure}

We further evaluated the geometric reconstruction performance by rendering depth maps in the optical demonstration, as shown in Fig.~\ref{fig:opt_depth}.
Here, positional shifts as well as scale and bias ambiguities in the depth maps were compensated by estimating alignment parameters using the Adam optimizer.  
Compared with the reference depth map in Fig.~\ref{fig:opt_depth}(a), DoF-NeRF in Fig.~\ref{fig:opt_depth}(c) exhibits more severe reconstruction artifacts than NeRF in Fig.~\ref{fig:opt_depth}(b) and EDoF-NeRF in Fig.~\ref{fig:opt_depth}(d).  
In contrast to the numerical demonstration, the qualitative difference between NeRF and EDoF-NeRF is less pronounced in the optical demonstration.  
However, the quantitative evaluation of the depth maps, summarized in Table~\ref{tab:opt_depth}, reveals higher reconstruction accuracy for EDoF-NeRF.  
The PSNR and SSIM values were calculated after linearly normalizing the depth maps using their global minimum and maximum values.
These quantitative results support the advantage of EDoF-NeRF over NeRF and DoF-NeRF.

\begin{table}[tbp]
  \centering
  \caption{Quantitative evaluation of the reconstructed novel-view images in the optical demonstration.}
  \label{tab:opt_rgb}
  \begin{tabular}{c | c c}
    \hline
    Evaluated model & PSNR~[dB]$\uparrow$ & SSIM$\uparrow$ \\
    \hline
    NeRF~(Fig.~\ref{fig:opt_rgb}(f)) & 24.6 & 0.742 \\
    DoF-NeRF~(Fig.~\ref{fig:opt_rgb}(g)) & 24.4 & 0.751 \\
    EDoF-NeRF~(Fig.~\ref{fig:opt_rgb}(h)) & \textbf{26.4} & \textbf{0.814} \\
    \hline
  \end{tabular}
\end{table}

\begin{table}[tbp]
  \centering
  \caption{Quantitative evaluation of the estimated depth maps in the optical demonstration.}
  \label{tab:opt_depth}
  \begin{tabular}{c | c c}
    \hline
    Evaluated model & PSNR~[dB]$\uparrow$ & SSIM$\uparrow$ \\
    \hline
    NeRF~(Fig.~\ref{fig:opt_depth}(b))      & 27.3 & 0.789 \\
    DoF-NeRF~(Fig.~\ref{fig:opt_depth}(c))  & 24.8 & 0.720 \\
    EDoF-NeRF~(Fig.~\ref{fig:opt_depth}(d)) & \textbf{28.0} & \textbf{0.795} \\
    \hline
  \end{tabular}
\end{table}

\section{Conclusion}

In this study, we proposed EDoF-NeRF, which incorporates a coded aperture into the NeRF rendering model to extend the DoF of synthesized novel views.
The coded PSF preserves high-frequency spatial information that is otherwise lost by conventional circular apertures under defocused conditions.
By jointly optimizing the scene representation and optical parameters through two-stage training, EDoF-NeRF recovers sharp, all-in-focus novel views from coded inputs.
Numerical and optical demonstrations confirmed that EDoF-NeRF outperforms NeRF and DoF-NeRF in terms of PSNR and SSIM for both reconstructed novel-view images and estimated depth maps.

The fidelity of EDoF reconstruction could be further improved by optimizing the coded-aperture pattern, which was empirically selected in this study, for example through end-to-end optimization~\cite{elmalem2018learned, dun2020learned, liu2021endtoend, pinilla2022hybrid, li2023optimizing}.
Moreover, because the proposed framework incorporates optical effects through the PSF, it is not limited to amplitude-modulation-based PSF engineering and can be naturally extended to phase-modulation-based approaches that can minimize light loss, such as wavefront coding~\cite{dowski1995extended}, rotating PSFs~\cite{pavani2008high, quirin2013depth}, and defocus-invariant phase masks~\cite{le2014optimized}.
We believe that EDoF-NeRF offers a promising direction for practical novel-view synthesis in virtual and augmented reality, dynamic scene capture, and intelligent vision systems.

\begin{backmatter}
\bmsection{Funding}~Japan Society for the Promotion of Science~(JP23K26567, JP23H05444, JP25K22227); JST-FOREST (JPMJFR2448); JST-ALCA-Next (JPMJAN25F1); Asahi Glass Foundation; SECOM Science and Technology Foundation.

\bmsection{Disclosures}
The authors declare no conflicts of interest.

\bmsection{Data Availability}
Data underlying the results presented in this paper are available from the corresponding author upon reasonable request.
\end{backmatter}

\bibliography{refs}

\begin{thebibliography}{10}
\newcommand{\enquote}[1]{``#1''}

\bibitem{mildenhall2021NeRF}
B.~Mildenhall, P.~P. Srinivasan, M.~Tancik, \emph{et~al.}, \enquote{{NeRF}:
  Representing scenes as neural radiance fields for view synthesis,}
  {\protect\JournalTitle{Communications of the ACM}} \textbf{65}, 99--106
  (2021).

\bibitem{li2022Rt}
C.~Li, S.~Li, Y.~Zhao, \emph{et~al.}, \enquote{{RT-NeRF}: Real-time on-device
  neural radiance fields towards immersive {AR/VR} rendering,} in
  \emph{Proceedings of the 41st IEEE/ACM International Conference on
  Computer-Aided Design,}  (2022), pp. 1--9.

\bibitem{zou2024ARthroNeRF}
X.~Zou, Z.~Zhang, A.~Schwarz, \emph{et~al.}, \enquote{{ARthroNeRF}: Field of
  view enhancement of arthroscopic surgeries using augmented reality and neural
  radiance fields,} in \emph{2024 IEEE International Symposium on Mixed and
  Augmented Reality (ISMAR),}  (IEEE, 2024), pp. 1197--1205.

\bibitem{croce2023Neural}
V.~Croce, G.~Caroti, L.~De~Luca, \emph{et~al.}, \enquote{Neural radiance fields
  ({NeRF}): Review and potential applications to digital cultural heritage,}
  {\protect\JournalTitle{The International Archives of the Photogrammetry,
  Remote Sensing and Spatial Information Sciences}} \textbf{48}, 453--460
  (2023).

\bibitem{mazzacca2023NeRF}
G.~Mazzacca, A.~Karami, S.~Rigon, \emph{et~al.}, \enquote{{NeRF} for heritage
  {3D} reconstruction,} {\protect\JournalTitle{International Archives of the
  Photogrammetry, Remote Sensing and Spatial Information Sciences}}
  \textbf{48}, 1051--1058 (2023).

\bibitem{cao2024Lightning}
J.~Cao, Z.~Li, N.~Wang, and C.~Ma, \enquote{Lightning {NeRF}: Efficient hybrid
  scene representation for autonomous driving,} in \emph{2024 IEEE
  International Conference on Robotics and Automation (ICRA),}  (IEEE, 2024),
  pp. 16803--16809.

\bibitem{shen2024Driveenv}
M.-Y. Shen, C.-C. Hsu, H.-Y. Hou, \emph{et~al.}, \enquote{{DriveEnv-NeRF}:
  Exploration of a {NeRF-based} autonomous driving environment for real-world
  performance validation,} {\protect\JournalTitle{ArXiv Preprint
  ArXiv:2403.15791}}  (2024).

\bibitem{ma2022Deblur}
L.~Ma, X.~Li, J.~Liao, \emph{et~al.}, \enquote{Deblur-{NeRF}: Neural radiance
  fields from blurry images,} in \emph{Proceedings of the IEEE/CVF Conference
  on Computer Vision and Pattern Recognition,}  (2022), pp. 12861--12870.

\bibitem{kaneko2022ar}
T.~Kaneko, \enquote{{AR-NeRF}: Unsupervised learning of depth and defocus
  effects from natural images with aperture rendering neural radiance fields,}
  in \emph{Proceedings of the IEEE/CVF Conference on Computer Vision and
  Pattern Recognition,}  (2022), pp. 18387--18397.

\bibitem{wu2022DoF}
Z.~Wu, X.~Li, J.~Peng, \emph{et~al.}, \enquote{{DoF-NeRF}: Depth-of-field meets
  neural radiance fields,} in \emph{Proceedings of the 30th ACM International
  Conference on Multimedia,}  (2022), pp. 1718--1729.

\bibitem{goodman2005Introduction}
J.~W. Goodman, \emph{Introduction to {Fourier} {Optics}} (Roberts and Company
  publishers, 2005).

\bibitem{gehm2007single}
M.~E. Gehm, R.~John, D.~J. Brady, \emph{et~al.}, \enquote{Single-shot
  compressive spectral imaging with a dual-disperser architecture,}
  {\protect\JournalTitle{Optics Express}} \textbf{15}, 14013--14027 (2007).

\bibitem{wagadarikar2009video}
A.~A. Wagadarikar, N.~P. Pitsianis, X.~Sun, and D.~J. Brady, \enquote{Video
  rate spectral imaging using a coded aperture snapshot spectral imager,}
  {\protect\JournalTitle{Optics Express}} \textbf{17}, 6368--6388 (2009).

\bibitem{llull2013coded}
P.~Llull, X.~Liao, X.~Yuan, \emph{et~al.}, \enquote{Coded aperture compressive
  temporal imaging,} {\protect\JournalTitle{Optics Express}} \textbf{21},
  10526--10545 (2013).

\bibitem{miyakawa2014coded}
R.~Miyakawa, R.~Mayer, A.~Wojdyla, \emph{et~al.}, \enquote{Coded aperture
  detector: an image sensor with sub 20-nm pixel resolution,}
  {\protect\JournalTitle{Optics Express}} \textbf{22}, 19803--19809 (2014).

\bibitem{bacca2020coupled}
J.~Bacca, L.~Galvis, and H.~Arguello, \enquote{Coupled deep learning coded
  aperture design for compressive image classification,}
  {\protect\JournalTitle{Optics Express}} \textbf{28}, 8528--8540 (2020).

\bibitem{ge2023coded}
Y.~Ge, G.~Qu, Y.~Huang, and D.~Liu, \enquote{Coded aperture compression
  temporal imaging based on a dual-mask and deep denoiser,}
  {\protect\JournalTitle{Journal of the Optical Society of America A}}
  \textbf{40}, 1468--1477 (2023).

\bibitem{wang2025dual}
G.~Wang, X.~Liu, and X.~Yuan, \enquote{Dual-optical-path coded aperture
  compressive temporal imaging,} {\protect\JournalTitle{Optics Letters}}
  \textbf{50}, 1865--1868 (2025).

\bibitem{ables1968Fourier}
J.~Ables, \enquote{Fourier transform photography: a new method for x-ray
  astronomy,} {\protect\JournalTitle{Publications of the Astronomical Society
  of Australia}} \textbf{1}, 172--173 (1968).

\bibitem{dicke1968Scatter}
R.~Dicke, \enquote{Scatter-hole cameras for x-rays and gamma rays,}
  {\protect\JournalTitle{Astrophysical Journal, vol. 153, p. L101}}
  \textbf{153}, L101 (1968).

\bibitem{fenimore1978Coded}
E.~E. Fenimore and T.~M. Cannon, \enquote{Coded aperture imaging with uniformly
  redundant arrays,} {\protect\JournalTitle{Applied Optics}} \textbf{17},
  337--347 (1978).

\bibitem{levin2007Image}
A.~Levin, R.~Fergus, F.~Durand, and W.~T. Freeman, \enquote{Image and depth
  from a conventional camera with a coded aperture,} {\protect\JournalTitle{ACM
  Transactions on Graphics (TOG)}} \textbf{26}, 70--es (2007).

\bibitem{zhou2009Coded}
C.~Zhou, S.~Lin, and S.~Nayar, \enquote{Coded aperture pairs for depth from
  defocus,} in \emph{2009 IEEE 12th International Conference on Computer
  Vision,}  (IEEE, 2009), pp. 325--332.

\bibitem{horisaki2020Deeply}
R.~Horisaki, Y.~Okamoto, and J.~Tanida, \enquote{Deeply coded aperture for
  lensless imaging,} {\protect\JournalTitle{Optics Letters}} \textbf{45},
  3131--3134 (2020).

\bibitem{silva2025toward}
J.~R. C. S.~A. Silva~Neto, H.~Kawachi, Y.~Yagi, and T.~Nakamura,
  \enquote{Toward all-in-focus lensless imaging with full-aperture radial
  masks,} {\protect\JournalTitle{Optics Express}} \textbf{33}, 48112--48129
  (2025).

\bibitem{kingma2014adam}
D.~P. Kingma and J.~Ba, \enquote{Adam: A method for stochastic optimization,}
  {\protect\JournalTitle{ArXiv Preprint ArXiv:1412.6980}}  (2014).

\bibitem{wang2004Image}
Z.~Wang, A.~C. Bovik, H.~R. Sheikh, and E.~P. Simoncelli, \enquote{Image
  quality assessment: from error visibility to structural similarity,}
  {\protect\JournalTitle{IEEE transactions on image processing}} \textbf{13},
  600--612 (2004).

\bibitem{schonberger2016Structure}
J.~L. Schonberger and J.-M. Frahm, \enquote{Structure-from-motion revisited,}
  in \emph{Proceedings of the IEEE Conference on Computer Vision and Pattern
  Recognition,}  (2016), pp. 4104--4113.

\bibitem{elmalem2018learned}
S.~Elmalem, R.~Giryes, and E.~Marom, \enquote{Learned phase coded aperture for
  the benefit of depth of field extension,} {\protect\JournalTitle{Optics
  Express}} \textbf{26}, 15316--15331 (2018).

\bibitem{dun2020learned}
X.~Dun, H.~Ikoma, G.~Wetzstein, \emph{et~al.}, \enquote{Learned rotationally
  symmetric diffractive achromat for full-spectrum computational imaging,}
  {\protect\JournalTitle{Optica}} \textbf{7}, 913--922 (2020).

\bibitem{liu2021endtoend}
Y.~Liu, C.~Zhang, T.~Kou, \emph{et~al.}, \enquote{End-to-end computational
  optics with a singlet lens for large depth-of-field imaging,}
  {\protect\JournalTitle{Optics Express}} \textbf{29}, 28530--28548 (2021).

\bibitem{pinilla2022hybrid}
S.~Pinilla, S.~R. Miri~Rostami, I.~Shevkunov, \emph{et~al.}, \enquote{Hybrid
  diffractive optics design via hardware-in-the-loop methodology for achromatic
  extended-depth-of-field imaging,} {\protect\JournalTitle{Optics Express}}
  \textbf{30}, 32633--32649 (2022).

\bibitem{li2023optimizing}
Y.~Li, Y.~Lyu, J.~Wang, \emph{et~al.}, \enquote{Optimizing wavefront coding for
  extended depth of field: a synchronous algorithm for optical element and
  decoding optimization,} {\protect\JournalTitle{Optics Letters}} \textbf{48},
  5847--5850 (2023).

\bibitem{dowski1995extended}
E.~R. Dowski~Jr and W.~T. Cathey, \enquote{Extended depth of field through
  wave-front coding,} {\protect\JournalTitle{Applied Optics}} \textbf{34},
  1859--1866 (1995).

\bibitem{pavani2008high}
S.~R.~P. Pavani and R.~Piestun, \enquote{High-efficiency rotating point spread
  functions,} {\protect\JournalTitle{Optics Express}} \textbf{16}, 3484--3489
  (2008).

\bibitem{quirin2013depth}
S.~Quirin and R.~Piestun, \enquote{Depth estimation and image recovery using
  broadband, incoherent illumination with engineered point spread functions,}
  {\protect\JournalTitle{Applied Optics}} \textbf{52}, A367--76 (2013).

\bibitem{le2014optimized}
V.~N. Le, S.~Chen, and Z.~Fan, \enquote{Optimized asymmetrical tangent phase
  mask to obtain defocus invariant modulation transfer function in incoherent
  imaging systems,} {\protect\JournalTitle{Optics Letters}} \textbf{39},
  2171--2174 (2014).

\end{thebibliography}

\end{document}